\documentclass[twocolumn]{aastex631}
\usepackage{placeins}

\usepackage{graphicx} 
\usepackage{comment}
\usepackage{amsmath}   
\usepackage{threeparttable}
\usepackage[utf8]{inputenc}

\begin{document}

\title{ATOMS-QUARKS survey: Inflow and infall in massive protocluster G318.049+00.086: Evidence of competitive accretion}

\author[0000-0002-8614-0025]{Shivani Gupta}
\affiliation{Indian Institute of Astrophysics, Koramangala II Block, Bangalore 560 034, India; \href{mailto:shivani.gupta@iiap.res.in}{shivani.gupta@iiap.res.in}}
\affiliation{Pondicherry University, R.V. Nagar, Kalapet, 605014, Puducherry, India}
\affiliation{S.N. Bose National Centre for Basic Sciences, Sector-III, Salt Lake, Kolkata 700106, India}

\author[0000-0003-0295-6586]{Tapas Baug}
\affiliation{S.N. Bose National Centre for Basic Sciences, Sector-III, Salt Lake, Kolkata 700106, India}

\author[0000-0002-6386-2906]{Archana Soam}
\affiliation{Indian Institute of Astrophysics, Koramangala II Block, Bangalore 560 034, India; \href{mailto:shivani.gupta@iiap.res.in}{shivani.gupta@iiap.res.in}}
\affiliation{Pondicherry University, R.V. Nagar, Kalapet, 605014, Puducherry, India}

\author[0000-0002-5286-2564]{Tie Liu}
\affiliation{Shanghai Astronomical Observatory, Chinese Academy of Sciences, 80 Nandan Road, Shanghai 200030, Peoples Republic of China}
\affiliation{Key Laboratory for Research in Galaxies and Cosmology, Shanghai Astronomical Observatory, Chinese Academy of Sciences, 80 Nandan Road, Shanghai 200030, Peoples Republic of China}

\author[0000-0001-5950-1932]{Fengwei Xu}
\affiliation{Max Planck Institute for Astronomy, Königstuhl 17, 69117 Heidelberg, Germany}

\author[0009-0007-9411-0284]{Satyajeet Moharana}
\affiliation{Korea Astronomy and Space Science Institute, 776 Daedeokdae-ro, Yuseong-gu, Daejeon 34055, Republic of Korea}
\affiliation{University of Science and Technology, 217 Gajeong-ro, Yuseong-gu, Daejeon 34113, Republic of Korea}

\author[0000-0003-1649-7958]{Guido Garay}
\affiliation{Chinese Academy of Sciences South America Center for Astronomy, National Astronomical Observatories, CAS, Beijing 100101, PR China}
\affiliation{Departamento de Astronomía, Universidad de Chile, Las Condes, 7591245 Santiago, Chile}

\author[0000-0002-3179-6334]{Chang Won Lee}
\affiliation{Korea Astronomy and Space Science Institute, 776 Daedeokdae-ro, Yuseong-gu, Daejeon 34055, Republic of Korea}
\affiliation{University of Science and Technology, 217 Gajeong-ro, Yuseong-gu, Daejeon 34113, Republic of Korea}

\author[0000-0002-9836-0279]{Siju Zhang}
\affiliation{Departamento de Astronom\'{i}a, Universidad de Chile, Camino el Observatorio 1515, Las Condes, Santiago, Chile}

\author[0009-0003-6633-525X] {Ariful Hoque}
\affiliation{S.N. Bose National Centre for Basic Sciences, Sector-III, Salt Lake, Kolkata 700106, India}

\author[0009-0002-6147-531X]{Puja Porel}
\affiliation{Indian Institute of Astrophysics, Koramangala II Block, Bangalore 560 034, India; \href{mailto:shivani.gupta@iiap.res.in}{shivani.gupta@iiap.res.in}}
\affiliation{Pondicherry University, R.V. Nagar, Kalapet, 605014, Puducherry, India}

\author{Lei Zhu}
\affiliation{Chinese Academy of Sciences South America Center for Astronomy, National Astronomical Observatories, CAS, Beijing 100101, PR China}

\author[0009-0004-6159-5375]{Dongting Yang}
\affiliation{School of Physics and Astronomy, Yunnan University, Kunming, 650091, People’s Republic of China}

\author[0000-0003-3343-9645]{HongLi Liu}
\affiliation{Department of Astronomy, Yunnan University, Kunming, 650091, China}

\author[0000-0001-9822-7817]{Wenyu Jiao}
\affiliation{Kavli Institute for Astronomy and Astrophysics, Peking University, Beijing 100871, People’s Republic of China}
\affiliation{Department of Astronomy, School of Physics, Peking University, Beijing, 100871, People’s Republic of China}

\author[0000-0001-8315-4248]{Xunchuan Liu}
\affiliation{Shanghai Astronomical Observatory, Chinese Academy of Sciences, 80 Nandan Road, Shanghai 200030, Peoples Republic of China}

\author[0000-0002-4719-3706]{Alik Panja}
\affiliation{Indian Institute of Astrophysics, Koramangala II Block, Bangalore 560 034, India; \href{mailto:shivani.gupta@iiap.res.in}{shivani.gupta@iiap.res.in}}

\author[0000-0001-7573-0145]{Xiaofeng Mai}
\affiliation{Shanghai Astronomical Observatory, Chinese Academy of Sciences, 80 Nandan Road, Shanghai 200030, Peoples Republic of China}

\author[0000-0001-7817-1975]{Yankun Zhang}
\affiliation{Shanghai Astronomical Observatory, Chinese Academy of Sciences, 80 Nandan Road, Shanghai 200030, Peoples Republic of China}

\author[0000-0001-9333-5608]{Shinyoung Kim}
\affiliation{Korea Astronomy and Space Science Institute, 776 Daedeokdae-ro, Yuseong-gu, Daejeon 34055, Republic of Korea}

\begin{abstract}
We present a gas kinematic study of the massive protocluster G318.049+00.086. The protocluster is reported to contain 12 prestellar core candidates and 4 protostellar cores. Filamentary structures are identified using the 1.3 mm dust continuum map, with four of them converge into a dense central region, forming a hub-filament system (HFS). High velocity gradients (10 - 20 km s$^{-1}$ pc$^{-1}$) derived from PV analysis of H$^{13}$CO$^{+}$ emission along three of those filaments are suggestive of mass inflow onto the central hub. A mass inflow rate higher than $10^{3}$ M$_{\odot}$ Myr$^{-1}$ along the filaments is indicating that the central hub is capable of forming massive star(s). Investigation of H$^{13}$CO$^{+}$ and CCH spectral profiles revealed the majority of the cores having the characteristic blue asymmetric line profiles, typical signature of gravitational collapse. The remaining few cores showed red asymmetric profiles, indicative of gas expansion. Also, the derived mass infall rates for the protostellar cores in hub-region is significantly higher in comparison to those located along the filaments. The mass-radius relationship of the cores revealed that the cores with red profiles reside in the massive star formation regime. However, the global velocity gradient along the filaments suggests that these particular cores are losing material to the hub. Our results are supporting a competitive accretion scenario of massive star formation where gas is expected to be funnelled from less gravitationally dominant cores to the cores located at the gravitationally favorable position.
\end{abstract}


\section{Introduction} \label{sec:intro}

Massive stars ($\geq$ 8 M$_{\odot}$) are crucial agents in galactic evolution owing to their strong radiative feedback, powerful stellar winds, and finally followed by an energetic supernova explosion toward the end of their lives. Yet, the formation mechanism of massive stars is still not full understood. Over the past decades, several theoretical models have been proposed to explain the formation of high-mass stars, though none have been conclusively validated observationally, primarily because of their rarity, rapid evolution, and their formation in deeply embedded clustered environments \citep{1987ARA&A..25...23S,2007ApJ...654..304K, 2007prpl.conf..149B,2014prpl.conf..149T, 2018ARA&A..56...41M}. 

Two basic classes of theory are under active study for the formation of massive stars -- (1) core accretion models or turbulent core model, and (2) clump-fed models, such as competitive-accretion model, inertial-inflow model, and global hierarchical collapse (GHC) model.

The turbulent-core model assumes a similar formation mechanism as of their lower-mass counterparts, that the formation of a massive star starts from the collapse of an isolated massive prestellar core, which is the gas reservoir for the final stellar mass \citep{2003ApJ...585..850M,2007ApJ...656..959K}. It assumes that stars of all masses form via disk-mediated accretion from dense isolated prestellar cores, except for massive stars need to overcome additional support mechanisms like radiation pressure and ionization feedback.
Prestellar cores having mass $>$16 M$_{\odot}$ within 0.01–0.1 pc of radius are considered as key evidence for the turbulent core model \citep{2025ApJS..280...33Y}. Notably, only a handful of studies have found the existence of massive prestellar cores ($\sim$20--30 M$_{\odot}$) that can lead to the formation of massive stars \citep[see][for example]{2013ApJ...779...96T, 2025A&A...696A..11V, 2025ApJS..280...33Y}. This low detection rate implies a very short lifetime for high-mass prestellar cores \citep{2018ARA&A..56...41M}. Factors, such as strong magnetic fields, high temperature, and concentrated initial density profiles are found to effectively contribute to reduce fragmentation within the massive prestellar core and thus lead to the formation of
more massive cores for high-mass star formation \citep{2021ApJ...912..159P,2025ApJ...980...87S,2025arXiv251115079Y}.

In contrast, the competitive accretion model is a theoretical framework proposed to explain how massive stars form within clustered environments \citep{1997MNRAS.285..201B,2001MNRAS.323..785B,2006MNRAS.370..488B,2007ARA&A..45..565M,2009MNRAS.400.1775S}. In this model, all stars — both low-mass and massive stars — form from a low-mass gravitationally bound prestellar core as a seed. These prestellar cores are embedded in a common massive clump or molecular cloud, where the bulk of the gas is not bound to individual cores but rather to the gravitational potential of the entire stellar cluster. It only accounts for mass accretion due to the gravity of the growing protostars (Bondi–Hoyle accretion), neglecting the pre-existing inflow at the larger scales.

The other sub-type of clump-fed model is the inertial-inflow model. According to this model, the prestellar cores that evolve into massive stars have a broad mass distribution but can accrete gas from pc scales reservoir through large-scale converging flows, with the pc-scale region around the prestellar core being turbulent and gravitationally unbound  \citep{2020ApJ...900...82P}. However, in contrary observations revealed most of the Galactic pc-scale massive clumps to be gravitationally bound irrespective of their evolutionary time-scale \citep{2016ApJ...829...59L, 2018MNRAS.473.1059U}. 

The GHC model \citep{2011MNRAS.411...65B,2017MNRAS.467.1313V} advocates a picture of molecular clouds in a state of hierarchical and chaotic gravitational collapse (multiscale infall motions), in which local centers of collapse develop throughout the cloud while the cloud itself is also contracting \citep{2019MNRAS.490.3061V}. This model predicts anisotropic gravitational contraction with longitudinal flow along filaments at all scales. It has been supported through observational evidences as well as numerical simulations \citep[see][and references therein]{2020ApJ...903...46C}.

All these ‘clump-fed’ models assume that massive stars born with low stellar masses but grow to much larger final stellar masses by accumulating mass from large-scale gas reservoirs beyond their natal dense cores. In these models, large-scale converging flows or global collapse of clumps are required to continuously feed mass into the dense cores. 

Recent observations revealed that massive star-forming regions frequently show evidence of large-scale inflow along filaments spanning 0.1 to several pc \citep{2013ApJ...766..115K, 2015ApJ...804..141Z}. This challenges the notion of isolated massive turbulent cores and supports a picture where hubs and cores build up mass dynamically via filamentary accretion.


Protoclusters provide a valuable glimpse into the early stages of clustered star formation, as they contain both prestellar and protostellar cores embedded within filamentary structures. These systems offer a unique opportunity to study large-scale inflows and small-scale infall processes simultaneously, which are essential for understanding how gas is channelled from pc scales down to individual cores \citep{2020A&A...642A..87K,2023ApJ...953...40Y}. In this paper, we studied the multi-scale gas dynamics of a massive protocluster (G318.049+00.086) that has reported evidence of the presence of multiple filaments and several prestellar and protostellar cores to understand the underlying star formation mechanism.

\subsection{Selected region}

The protocluster G318.049+00.086, also known as I14498-5856, is located at a distance of 2.90$^{+0.52}_{-0.47}$ kpc \citep{2019ApJ...885..131R} and has a local standard of rest velocity (V$_{\rm{lsr}}$) of -49.82 km s$^{-1}$ \citep{2021RAA....21...14Y}. The mass and dust temperature ($T_{\rm{dust}}$) of the protocluster are 1023.3 M$_{\odot}$ and 26.7 K, respectively \citep{2018MNRAS.473.1059U}. Using the CO (4--3) line, \citet{2021RAA....21...14Y} also reported that the global infall velocity and mass infall rate for this cluster to be V$_{\rm{in}}$ = 3.77 $\pm$ 0.07 km s$^{-1}$ and $\dot{M}_{\rm{in}}$ = (1.56 $\pm$ 0.03) $\times$ 10$^{4}$ M$_{\odot}$ Myr$^{-1}$.

\begin{figure*}[!htb]
    \centering
    \includegraphics[scale=0.45]{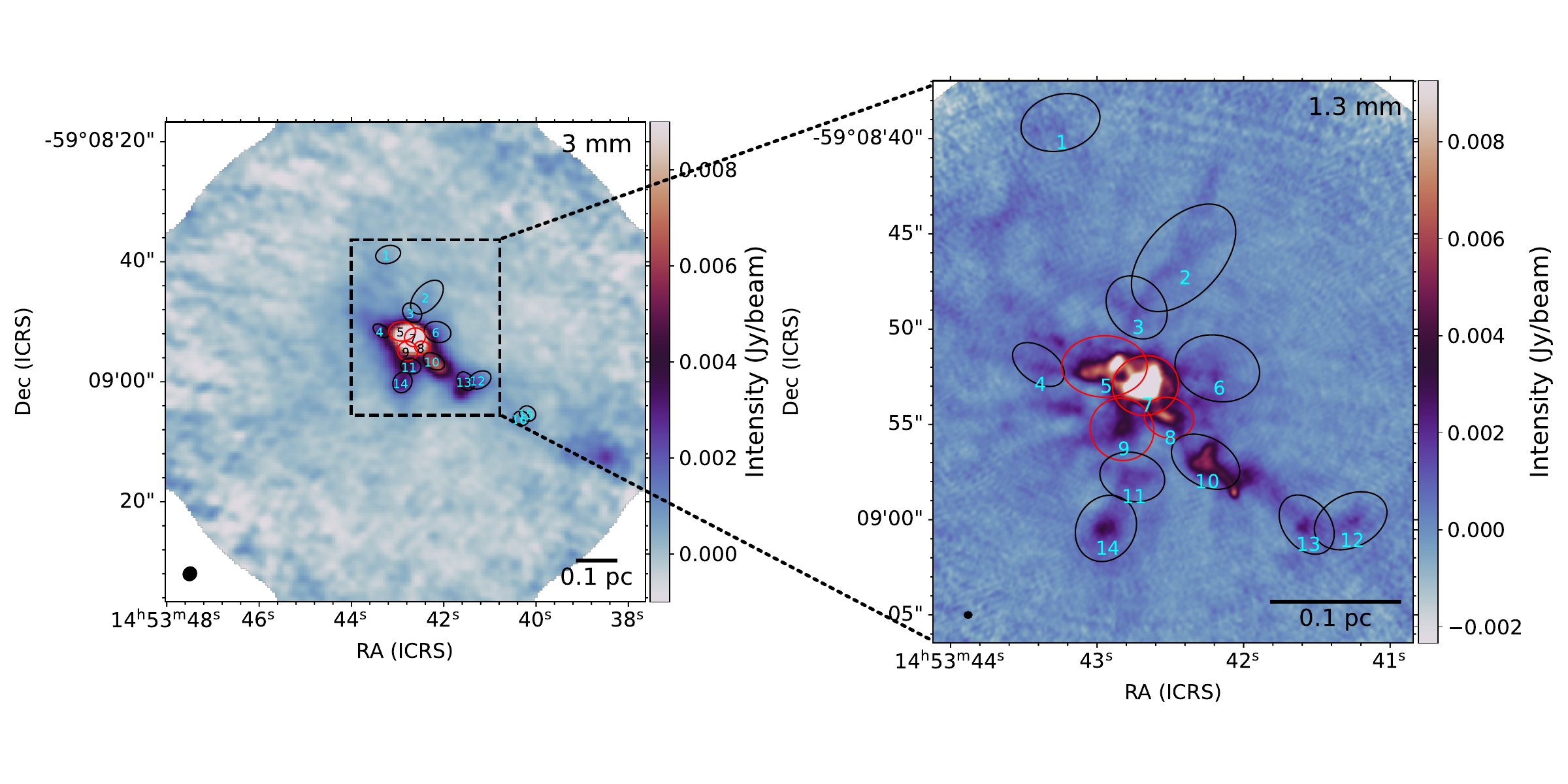}
    \caption{The {continuum map} of the protocluster at 3 mm (left panel) from the ATOMS survey and 1.3 mm (right panel) from the QUARKS survey. The overlaid ellipses {denote the} dense cores (red: protostellar {cores}, black: prestellar {core candidates}) identified at 870 $\mu$m by \cite{2024ApJS..270....9X}. The beam size and a scale bar are added at the bottom left and right corners of each panel.}
    \label{fig:band6}
\end{figure*}

Recently, \cite{2024ApJS..270....9X} identified 16 dense cores in the protocluster using the 870 $\mu$m dust continuum map observed using the Atacama Large Millimeter/submillimeter Array (ALMA; see their Figure 2b) and also derived several physical parameters (such as mass, radius, temperature, and surface density) of all the cores. Further, they also classified these cores as prestellar core candidates or protostellar {cores} based on their association with the bipolar outflows traced in the CO (3--2) line by \citet{2020ApJ...890...44B}. In total, they reported the presence of 4 protostellar {cores} and 12 prestellar {core candidates} in this protocluster. We adopted these cores in our analyses. 
In this paper, we analyzed the filamentary structures traced by the 1.3 mm dust continuum and velocity distribution along their skeletons using H$^{13}$CO$^{+}$ (1--0), as well as infalling motions towards the embedded dense cores using H$^{13}$CO$^{+}$ (1--0) and CCH (J=3/2--1/2) F(2--1) to look for the observational evidence for the massive star formation theory.
Figure \ref{fig:band6} represents the 3 mm (left panel) and 1.3 mm (right panel) dust continuum maps of the protocluster obtained from the ATOMS and QUARKS surveys, respectively (see Section \ref{sec:obs} for the surveys). The adopted protostellar cores and prestellar core candidates are represented as red and black ellipses, respectively.

Note that cores 15 and 16 are not present within the Field of View (FOV) of the 1.3 mm QUARKS data. Hence, we have not considered these cores in our subsequent analysis. Moreover, cores 1, 2, 3, and 4 are also excluded from further analysis. Core 1 is located significantly away from the main filamentary structure identified in the 1.3 mm QUARKS continuum map (Section \ref{sec:filament}), while cores 2, 3, and 4 appear to be tracing elongated filamentary material rather than distinct compact cores. This difference likely arises from the varying spatial filtering and sensitivity between the datasets, as the ASSEMBLE data include only the 12 m array, whereas the QUARKS data combine the 12 m and 7 m arrays, enabling better recovery of extended emission.

The structure of the paper is as follows. In Section~\ref{sec:obs}, we present the observations and data used in this study. Section~\ref{sec:result} elaborates the analysis and results obtained from the observed data. In Section~\ref{sec:discussion}, we discuss the overall gas dynamics and the star formation scenario in the region. Finally, we summarize our results in Section~\ref{sec:summary}.

\begin{table*}
\centering
    \caption{{Main targeted molecular lines in the study}}
    \scriptsize
    \label{tab:lines}
    \begin{tabular}{lcccccc} 
        \hline
        Species & Transition & Rest Frequency   & $\delta v$ & Beam Size &  rms  & Spatial scale\\ 
        &    & (GHz)& (km s$^{-1}$) & ($\prime\prime$) & (mJy beam$^{-1}$) &  (pc$^2$)  \\ 
        \noalign{\vskip 2pt}
        \hline
        \noalign{\vskip 2pt}
        \multicolumn{7}{l}{{ATOMS survey -- ALMA Band 3 (12m + ACA Combined})}\\
        \noalign{\vskip 2pt}
        \hline
        3 mm continuum & $\cdots$ & $\cdots$ & $\cdots$ & 2.34 $\times$ 2.16 & $\cdots$ &  0.033 $\times$ 0.030\\
         H$^{13}$CN & (1--0)& 86.339918  &  0.424  & 2.34 $\times$ 2.16  & 11.3 &  0.033 $\times$ 0.030 \\
          H$^{13}$CO$^{+}$ & (1--0) & 86.754288  &  0.422  & 2.66 $\times$ 2.52 & 11.1 &  0.037 $\times$ 0.035\\
          CCH & (1--0) &87.316898 &  0.419  & 2.63 $\times$ 2.50 & 11.3 &  0.037 $\times$ 0.035\\ 
          CS & (2--1)  &97.980953 &  2.973  & 2.36 $\times$  2.18 & 5.0 & 0.033 $\times$ 0.031\\ 
          \hline
          \noalign{\vskip 2pt}
          \multicolumn{7}{l}{{ASSEMBLE survey -- ALMA Band 6 (12m only)}}\\
          \noalign{\vskip 2pt}
          \hline
          CH$_3$OH &(13$_{1,12}$--13$_{0,13}$) & 342.729781 & 0.986   & 1.11 $\times$ 0.82 & 6.5 & 0.016 $\times$ 0.012\\ 
          H$^{13}$CN & (4--3) & 345.339760 & 0.980  & 1.08 $\times$ 0.80 & 6.3 & 0.015 $\times$ 0.011 \\  
        \noalign{\vskip 1pt}
        \hline 
        \noalign{\vskip 2pt}
        \multicolumn{7}{l}{{QUARKS survey -- ALMA Band 6 (12m + ACA Combined)}}\\
        \noalign{\vskip 2pt}
        \hline
        1.3 mm continuum & $\cdots$ & $\cdots$ & $\cdots$ & 0.40 $\times$ 0.34 & $\cdots$ &  0.0056 $\times$ 0.0048\\
        \hline
    \end{tabular}
\end{table*}

\section{Observation and Data Reduction} \label{sec:obs}
\subsection{ATOMS survey}
ALMA data for the protocluster was obtained as a part of the large survey named the ALMA three-millimeter observations of massive star-forming regions survey (ATOMS; Project ID: 2019.1.00685.S; PI: Tie Liu). Calibration of the 12-m array data and ACA data were done separately using the Common Astronomy Software Applications (CASA) package version 5.6 \citep{mcmullin2007asp}, and subsequently the visibility data were combined and re-imaged to recover the extended emission. The maximum recoverable scale (MRS) of 12m + ACA combine data reaches $\sim$ 87$^{\prime\prime}$ \citep{2023MNRAS.520.3259X}. Further details about the survey and data reduction can be found in \cite{2020MNRAS.496.2790L}. All the images used in our analysis are primary beam corrected. The 3 mm ($\sim$99.93 GHz) dust continuum map having a beam size of 2.34$^{\prime\prime}$ $\times$ 2.16$^{\prime\prime}$ is constructed using data from line-free spectral channels. {In this study, we utilized the H$^{13}$CO$^{+}$ (1--0), CCH F(2--1), and CS (2--1) transitions, with spectral resolutions of 0.422 km s$^{-1}$, 0.419 km s$^{-1}$, and 2.973 km s$^{-1}$, respectively.} The velocity-integrated intensity maps of all the molecules observed in the ATOMS survey are shown in Figure \ref{fig:moment0_all} (Appendix \ref{moment_map}).

\subsection{ASSEMBLE survey}
The G318.049+00.086 protocluster was also observed as part of the ALMA Survey of Star Formation and Evolution in Massive Protoclusters with Blue Profiles (ASSEMBLE; Project ID: 2017.1.00545.S; PI: Tie Liu). The  MRS of this data is $\sim$ 8.45$^{\prime\prime}$ \citep{2023MNRAS.520.3259X}. High-density tracers such as H$^{13}$CN (4–3) and hot-core molecular lines like CH$_3$OH (13$_{1,12}$–13$_{0,13}$) were employed to determine the core systemic velocity, with a spectral resolution of 0.98 km s$^{-1}$ \citep{2023MNRAS.520.3259X,2024ApJS..270....9X}.

\subsection{QUARKS survey}
Furthermore, G318.049+00.086 was observed as part of the Querying Underlying mechanisms of massive star formation with ALMA-Resolved gas Kinematics and Structures (QUARKS) survey, complementing the ATOMS survey with an improved angular resolution ($\sim$ 0.3$^{\prime\prime}$). The corresponding 1.3 mm dust continuum map is prepared from the available line-free spectral channels \citep{2024RAA....24b5009L}. QUARKS also have ACA observations (\citet{2024RAA....24f5011X}, Quarks paper II). The combine data has MRS of $\sim$ 8.45$^{\prime\prime}$ \citep{2024RAA....24f5011X}. The ACA observations help to cover the missing flux up to 30$^{\prime\prime}$, which is important to the science goals.\\

Table \ref{tab:lines} summaries all the lines and continuum data used in this study, along with their observational parameters.

\section{Result} \label{sec:result}
In this section, we present the analyses and results of the observed data. Filaments were already identified in the region by using H$^{13}$CO$^{+}$ in a previous study by \citet{2022MNRAS.514.6038Z}. In the following section, we elaborate on retracing the filament using H$^{13}$CO$^{+}$  and also on identification of the filaments by using 1.3 mm dust continuum in the region.

\subsection{Identification of filaments}\label{sec:filament}


\begin{figure}
    \centering
    \includegraphics[scale=0.51]{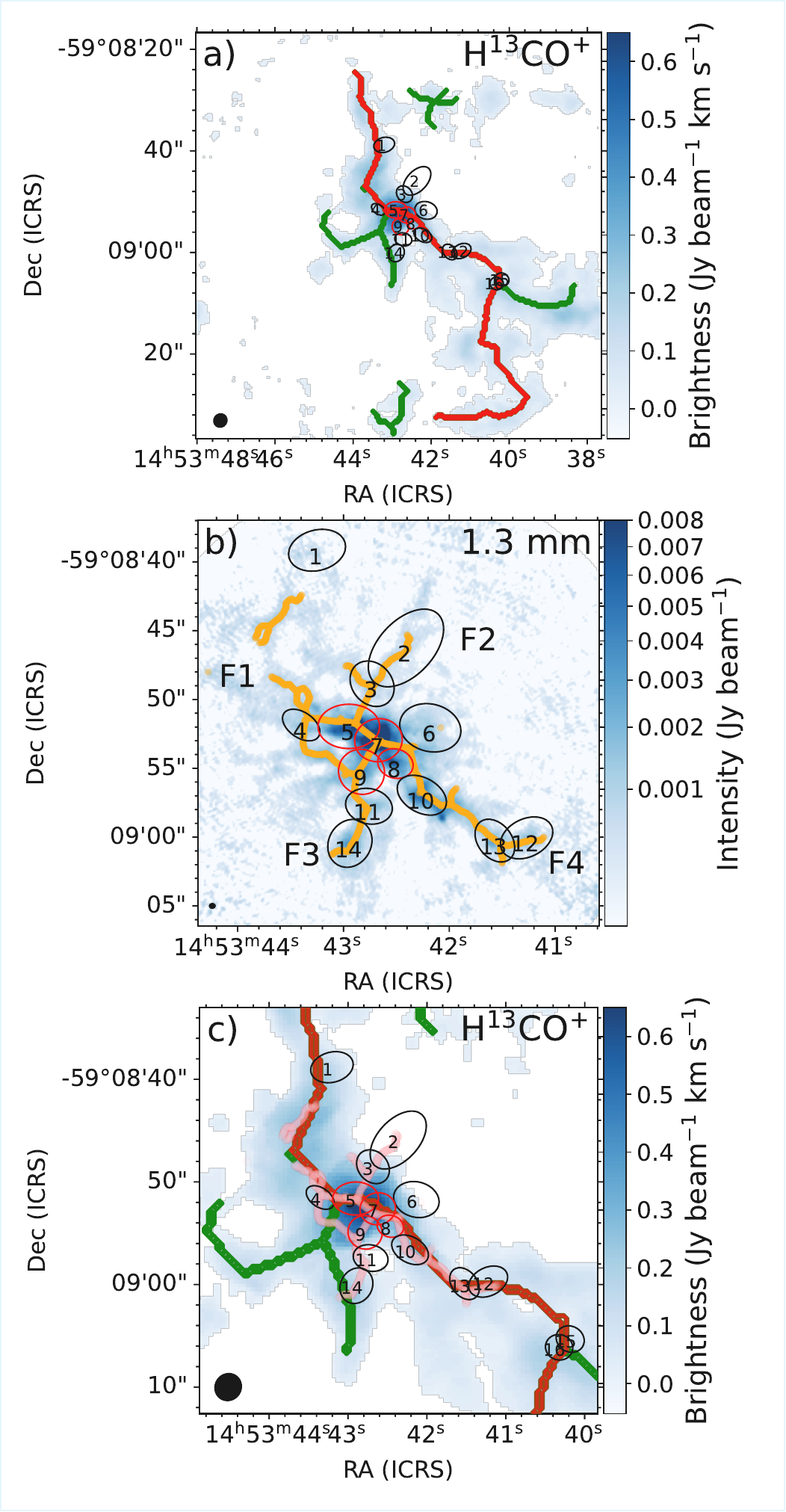}
    \caption{(a) The identified H$^{13}$CO$^{+}$ filaments overlaid on the moment 0 map of H$^{13}$CO$^{+}$ above the $5\sigma$ level. The red colored filament marks the longest filament and the green shows other filaments present in the region. (b) Displays the filaments (orange lines) identified in the 1.3 mm continuum emission overlaid on the 1.3 mm dust continuum map of the protocluster. The primary filaments are also labeled as F1, F2, F3, and F4. (c) Overlay of H$^{13}$CO$^{+}$  and 1.3 mm  filaments on H$^{13}$CO$^{+}$ moment 0. In all the panels, the identified cores by \citet{2024ApJS..270....9X} are shown and labeled.}
    \label{fig:filament}
\end{figure}


\cite{2022MNRAS.514.6038Z}, as part of the ATOMS survey, identified filamentary structures in this protocluster by applying the Python-based \texttt{FilFinder} algorithm  \citep{2015MNRAS.452.3435K} on the moment-0 map of H$^{13}$CO$^{+}$ emission. In our analysis, the previously identified filamentary structures were well-reproduced with the same \texttt{FilFinder} algorithm. The identified filaments are shown in Figure \ref{fig:filament}a. In addition to the main filament spine, \texttt{FilFinder} also identifies extended branches, shown in green, while the longest filament is highlighted in red (see Figure \ref{fig:filament}a).


We also employed the \texttt{FilFinder} algorithm to the 1.3 mm continuum map obtained in the QUARKS survey, as the continuum emission has good spatial resolution. The identified filaments are shown in Figure \ref{fig:filament}b. The filaments were also labeled as F1, F2, F3, and F4. Positions of the cores identified by \cite{2024ApJS..270....9X} are also marked in the figure. As can be seen in the figure, 12 out of the 14 identified cores are associated with these filaments.

In Figure \ref{fig:filament}c, we overlay the filamentary structures obtained from H$^{13}$CO$^{+}$ and 1.3 mm  continuum emission to examine their spatial correlation. We note that the longest filaments obtained from both the emissions strongly coincide with each other. This enables us for a comparative structural and dynamical analysis of the gas within the filaments.



\subsection{Column density map}
For better understanding of the distribution of the gas, we constructed the column density map of the protocluster. To generate the map, we chose the H$^{13}$CO$^{+}$ line, and it was derived using the formula obtained under the assumption of local thermodynamic equilibrium and the emission is optically thin \citep[LTE;][]{1991ApJ...374..540G},
\begin{align} \label{col_den}
N = & \, \frac{3k_{\rm B}}{8 \pi^3 B \mu^2} \cdot \frac{ \left( T_{\rm ex} + \frac{hB}{3k_{\rm B}} \right) }{(J + 1)} \cdot 
\frac{ \exp \left( \frac{hBJ(J+1)}{k_{\rm B} T_{\rm ex}} \right) }{1 - \exp \left( \frac{-h \nu}{k_{\rm B} T_{\rm ex}} \right)} \notag \\
& \times \frac{1}{J(T_{\rm ex}) - J(T_{\rm bg})} \int T_{\rm mb} \: dv,
\end{align}
where $B$ and $\mu$ are the rotational constant and permanent dipole moment of the molecule, respectively, $J$ is the rotational quantum number of the lower state in the observed transition, $k_{\rm B}$ is the Boltzmann constant, $\nu$ is the frequency and $h$ is the Planck constant. $T_{\rm bg}$ = 2.73 K is the cosmic microwave background temperature, $T_{\rm mb}$ is the main beam brightness temperature, and $T_{\rm ex}$ is the excitation temperature. $J(T) = T_{0}/(e^{T_{0}/T}-1)$ represents the source function, where $T_{0}$ = $h\nu/k_{\rm B}$. For this H$^{13}$CO$^{+}$ (1--0) transition, we used $J$ = 0, $\mu$ = 3.89 D, and $B$ = 43.377302 GHz \citep{FT9938902219, 1994JChPh.101.8945Y, 2007ApJ...662..771L,2012ApJ...756...60S}. With all these values Equation \ref{col_den} is simplified to,

\begin{align}
N({\rm H^{13}CO^{+}}) = & \, 2.43 \times 10^{11} \cdot (T_{\rm ex} + 0.69) \cdot \frac{1}{1 - \exp \left( \frac{-4.16}{T_{\rm ex}} \right)} \nonumber \\
& \times \left( \frac{4.16}{\exp \left( \frac{4.16}{T_{\rm ex}} \right) - 1} - 1.16 \right)^{-1} \int T_{\rm mb}\:dv
\end{align}
 Under the assumption of LTE, all levels are populated according to the same excitation temperature ($T_{\rm ex}$). In this calculation, we assume that $T_{\rm ex}$ is equal to the dust temperature of the clump ($T_{\rm dust}$). $\int T_{\rm mb}\:dv$ is the integrated intensity of the H$^{13}$CO$^{+}$ emission. For the construction of the column density map ($N({\rm H^{13}CO^{+}})$), we considered only the pixels with integrated intensity $>5\sigma$ in the moment-0 map. The remaining pixels were masked out from the final map.

 \begin{figure}[!htb]
    \centering
    \includegraphics[scale=0.32]{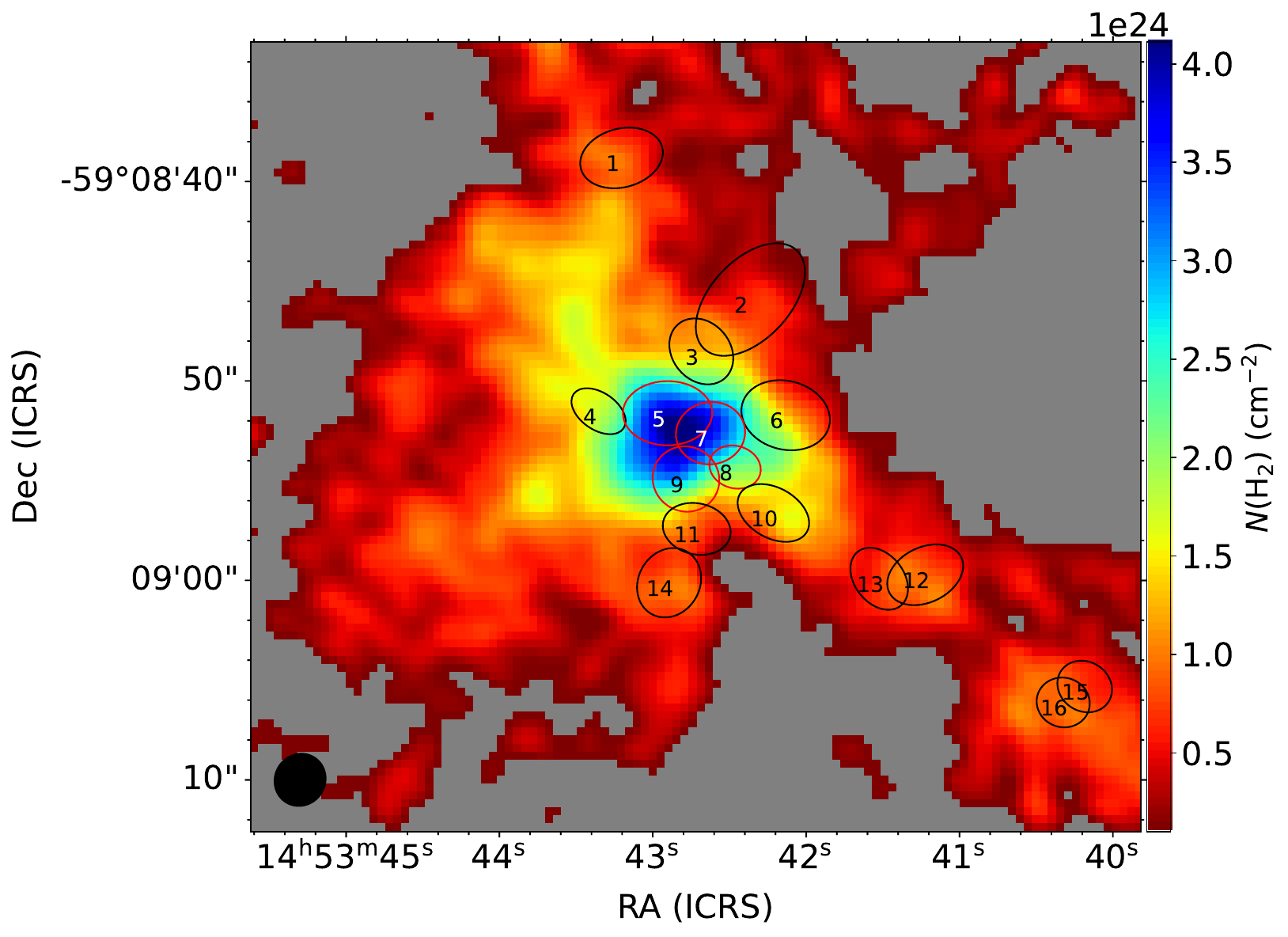}
    \caption{$N(\rm H_{2})$ column density map of the observed region derived using the H$^{13}$CO$^{+}$ line emission {above the $3\sigma$ level}. Positions of the cores identified by \cite{2024ApJS..270....9X} are also marked in the figure. }
    \label{fig:nh2}
\end{figure}

Conversion of $N({\rm H^{13}CO^{+}})$ column density map into hydrogen column density ($N(\rm H_{2})$) requires the [H$^{13}$CO$^{+}$/H$_2$] abundance ratio. However, we found that the corresponding abundance varies from 10$^{-10}$ to 10$^{-12}$ depending on various physical conditions \citep[e.g.,][and references therein]{1987ApJ...315..621B, 1987ApJ...313..320F, 2000A&A...355..499G, 2006ApJ...636..261H, 2021A&A...648A.120R,2024MNRAS.535.1364Z}. Thus, we derived the $N(\rm H_{2})$ from $N({\rm H^{13}CO^{+}})$ using the mean of the reported abundance ratio of 10$^{-11}$. The final column density map of the region is presented in Figure~\ref{fig:nh2}. Note that the $N$(H$_2$) ranges between (0.5-4.0)$\times$10$^{24}$ cm$^{-2}$ with the highest being at the hub region. 

The derived $N$(H$_2$) value represents a lower limit to the true column density, as the assumption of the line to be optically thin is not always valid, specifically in the dense region (i.e., dense cores, discussed in later sections). 


{If we consider H$^{13}$CO$^{+}$ has a moderate opacity ($\tau$ $\sim$ 0.5-1.5), the optically thin assumption would underestimate the column density by a factor $\tau/(1-e^{-\tau})$, corresponding to an increase of approximately $\sim$ 27\%--93\% relative to the optically thin estimation. This translates directly into a similar increase in the derived filament mass.
Further, we assumed a constant dust temperature ($T_{\rm dust}$) across the region. A plausible uncertainty of $\pm 10$ K to our adopted value introduces an uncertainty of approximately 33\% to 36\% in the derived $N(\mathrm{H}_2)$. Specifically, increasing the temperature by 10 K results in a $\sim$33\% increase in the column density (and mass), while decreasing the temperature by 10 K leads to a $\sim$36\% decrease. Finally, the conversion from $N(\mathrm{H^{13}CO^+})$ to $N(\rm{H}_2)$ depends critically on the assumed abundance ratio. While we adopted a representative value of $[\rm{H^{13}CO^+}/\rm{H}_2]=10^{-11}$, reported values span a wide range from $10^{-10}$ to $10^{-12}$. Adopting these extreme values would decrease or increase the inferred column density by an order of magnitude, respectively, implying that the total mass estimates could vary by up to a factor of 10.}

\subsection{Dynamical properties of filaments} \label{sec:vel_grad}

\begin{figure}
\centering
\includegraphics[width=\linewidth]{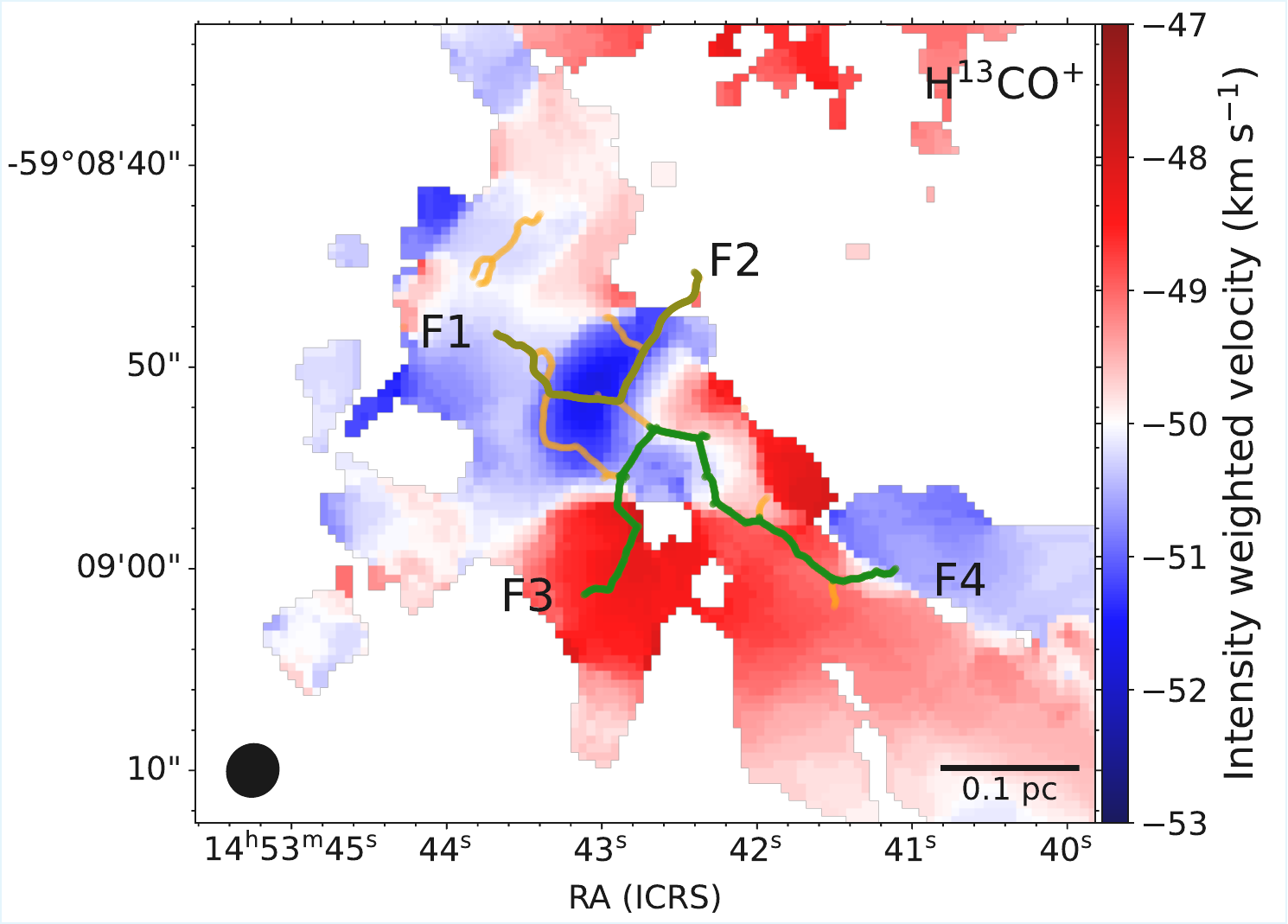}
\caption{Filaments identified in the 1.3 mm continuum emission overlaid on the moment 1 map of H$^{13}$CO$^{+}$ (1--0). The olive and green colors mark the primary filament paths (i.e., F1-F2 and F3-F4, respectively) used to derive the PV diagrams.}
\label{fig:moment1}
\end{figure}

\begin{figure}
\centering
\includegraphics[width=\linewidth]{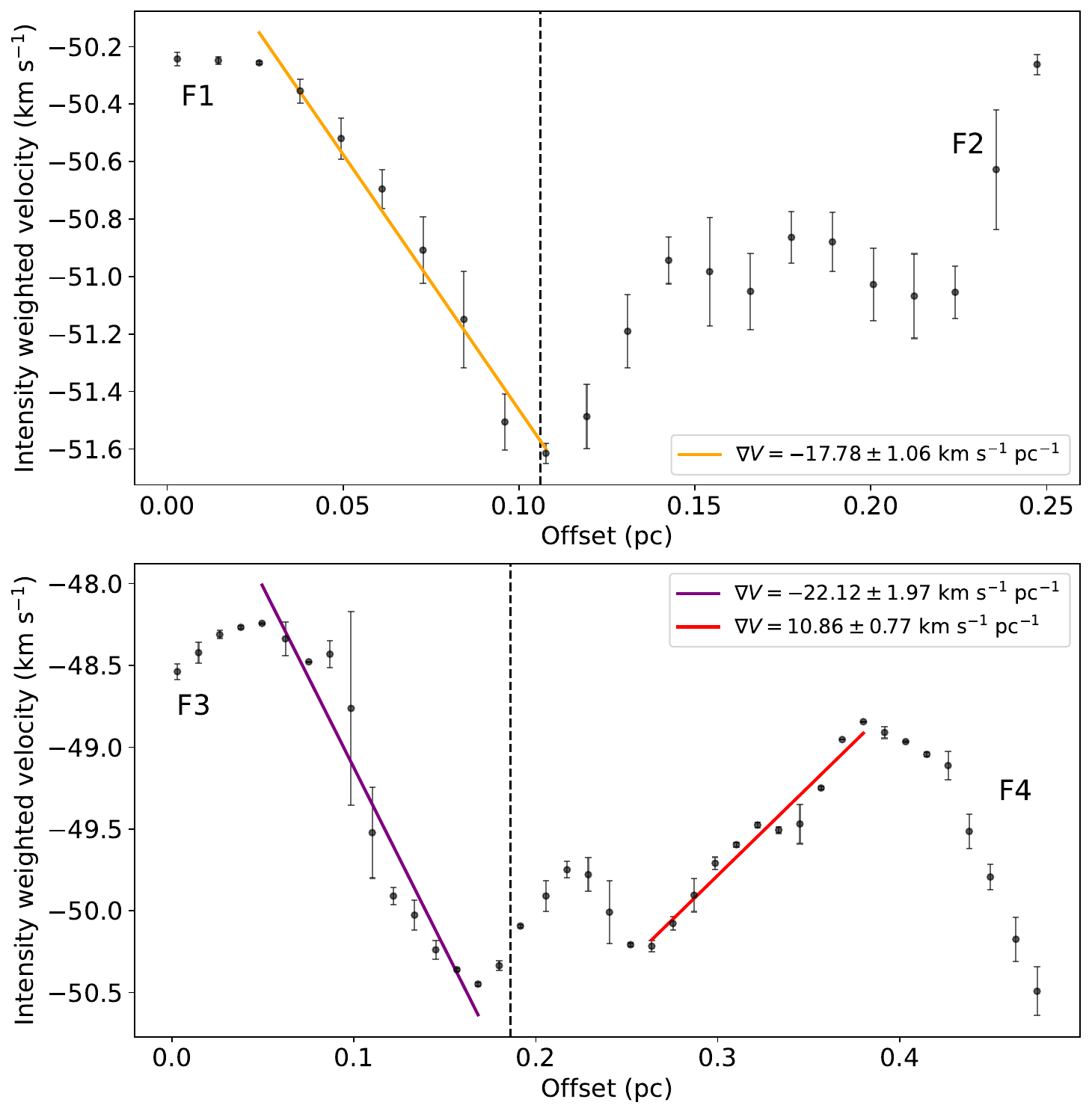}
\caption{The upper panel shows the PV diagram along the filament path F1 and F2, and the lower panel shows the PV along F3 and F4. The PV diagrams are uniformly sampled along the filament skeletons with a step size of 0.016 pc, corresponding to half of the synthesized beam size at the source distance. The velocity error bars represent the standard deviation of the velocities within each sampled bin along the offset. The vertical-dashed line in each panel marks the position of the hub. The orange, purple and red lines indicate the least squares fit to F1, F3 and F4 segments of the PV diagrams in both the panels, respectively.}
\label{fig:vel_grad}
\end{figure}

Figure \ref{fig:moment1} displays the moment 1 map (intensity weighted velocity map) of H$^{13}$CO$^{+}$, overlaid with the filamentary structure from the 1.3 mm continuum, providing insight into the underlying velocity distribution.
It has been reported in the previous studies that the filaments play an important role in transporting gas and aiding in star formation. To assess for such possibility, we examined the position-velocity (PV) diagram of the filaments {\citep{2022ApJ...940..112K}}. The PV diagrams of the filaments were constructed by determining the centroid velocity along each of these filament skeletons. Note that the four filaments labeled as F1, F2, F3, and F4 converge toward the central hub region. For convenience we combined them in two groups, i.e., F1\&F2 and F3\&F4 for construction of the PV diagrams and also to study the kinematics of the inflowing gas (see olive skeleton for F1\&F2 and green skeleton for F3\&F4 in Figure \ref{fig:moment1}). The corresponding PV diagrams are presented in Figure~\ref{fig:vel_grad}. The starting point for both the PV diagrams were defined at the eastern end of filaments F1 and F3, respectively. It can be seen in both the PV diagrams, a global minimum is present towards the location of the hub, suggesting for an inflow of gas along the filaments toward the hub. This inflow may facilitate conditions conducive to the formation of massive stars.

For estimation of the velocity gradients, we performed least-square fits to different parts of the PV diagrams signifying the filaments (see Figure~\ref{fig:vel_grad}). The slopes of these fits represent the mean velocity gradients. The estimated velocity gradients along filaments F1, F3, and F4 are 22.5 km s$^{-1}$ pc$^{-1}$, 22.1 km s$^{-1}$ pc$^{-1}$, and 10.9 km s$^{-1}$ pc$^{-1}$, respectively. The uncertainties for these gradients are obtained from the error associated with the least-squares fit (listed in Table~\ref{tab:mass_aacretion}). Note that we were unable to determine a consistent velocity gradient along the filament F2 due to significant velocity fluctuations along this filament. \citet{2022MNRAS.514.6038Z} also found velocity gradients using the H$^{13}$CO$^{+}$ along the largest path in the hub region, following the filamentary structure from F1 to F4.

Such a high velocity gradients within very short distance ($\sim$ 0.1-0.2 pc), indicates the dominance of gravity on inflow and accumulation of gas in dense cores or hub regions. In fact, \citet{2022MNRAS.514.6038Z} noted that in other protoclusters in ATOMS sample the velocity gradients also increase drastically within shorter separations ($\sim$ 10-40 km s$^{-1}$ pc$^{-1}$) compared to those derived for larger separations ($\sim$ 1 km s$^{-1}$ pc$^{-1}$ at 1.5 pc). 

\subsection{Mass inflow rate along the filaments}
For estimation of the mass flow along the filaments, we first derived the mass of the filament segments that are showing the velocity gradients. The filament mass was estimated from the H$_{2}$ column density map.
The calculated mass ($M_{\rm fil}$) and line mass ($m_{\rm fil}$) of the filaments are listed in Table~\ref{tab:mass_aacretion}. The uncertainty in $M_{\rm fil}$ is derived by adopting the uncertainty in distance to the protocluster (0.5 kpc) and the standard deviation in H$_{2}$ column density within the filament segment. The estimated filament masses are typical for Galactic filaments of the order of 100 M$_\odot$. However, the estimated line mass of the filaments are found to be 992-1746 M$_\odot$ pc$^{-1}$ which are typical for giant filaments \citep[see][for details]{2023ASPC..534..153H}. Such a high concentration of gas in these filaments are possibly due to their association with the massive protocluster.

To quantify the mass flow along the filaments, we followed a similar methodology described in \citet{2013ApJ...766..115K}. Assuming a simple cylindrical model, the equation for mass inflow rate of the axial flow motion can be written as
\begin{equation}
\dot{M}_{\parallel}=\frac{\nabla V_{\parallel}M_{\rm{fil}}}{\tan\alpha}
\end{equation}
where $\nabla$$V_\parallel$ is the velocity gradient measured along the filament, $M_{\rm fil}$ is the mass of the filament segment, and $\alpha$ is the angle between the axis of the cylinder and plane of the sky. The mass inflow rates along the filaments F1, F3 and F4 were computed using the estimated filament masses and the velocity gradients (see Table~\ref{tab:mass_aacretion}), and assuming the inclination angle ($\alpha$) of 45$^{\circ}$ and are found to be
in the order of 10$^3$ M$_{\odot}$ Myr$^{-1}$ (see also Table~\ref{tab:mass_aacretion} for exact values). Two extreme cases for $\alpha$, i.e., 0$^{\circ}$ and 90$^{\circ}$ signify the non-existence of the observed velocity gradients along filaments and non-existence of the filaments, respectively. However, such possibilities can be excluded from the fact that the filamentary structures do exist and the velocity variation along them is also significantly measurable. We thus assume $\tan$ $\alpha$ = 1 in our calculation, and the estimated velocity gradients and mass inflow rates are expected to be uncertain by a factor of 2 when the $\alpha$ changes between 30$^{\circ}$ and 60$^{\circ}$ {\citep{2025A&A...704A..64W}}. 

\begin{table*}
\begin{center}
    \caption{Derived parameter of the filaments, {inflow rates and mass transfer of the filaments feeding the hub}}
    \label{tab:mass_aacretion}
    \begin{tabular}{lcccccccc} 
        \hline
        Filament  & $L_{\rm fil}$ & $M_{\rm fil}$ & $m_{\rm fil}$ & $\nabla$$V_{\parallel}$ & $\dot{M}_{\parallel}$   & $\dot{M}_{\parallel,\rm line}$ & $\Delta\dot{M}_{\parallel}$ & $\Delta M$ \\ 
           ID &  (pc)         & (M$_{\odot}$) & (M$_{\odot}$  pc$^{-1}$)  &  (km s$^{-1}$ pc$^{-1}$)           &(M$_{\odot}$ Myr$^{-1}$) & (M$_{\odot}$ Myr$^{-1}$pc$^{-1}$)  & {(M$_{\odot}$ Myr$^{-1}$)} & {(M$_{\odot}$)} \\ 
        (1) &   (2)         &       (3)     &       (4)     &           (5)      &     (6)    & (7) &  {(8)}   &   {(9)}  \\
         \noalign{\vskip 3pt}
        \hline
        \noalign{\vskip 2pt}
        F1  & 0.07& 99$\pm$33  & 1414$\pm$473 & {17.8}$\pm${1.1} & {1762}$\pm${598} & {25170}$\pm${8540} &  {352$\pm$120}  &  {7.0$\pm$2.4} \\
        F3  & 0.13 & 227$\pm$76 & 1746$\pm$585 & 22.1$\pm${2.0} & {5017}$\pm${1735}  & {41808}$\pm$ {14458} & {1003$\pm$347}  & {20.0$\pm$7.0} \\
        F4  & 0.12 & 119$\pm$40 & 992$\pm$333 & {10.9}$\pm${0.8} & {1297}$\pm$445  & {10808}$\pm${3708} &  {259$\pm$89}  &  {5.2$\pm$1.8} \\
        \noalign{\vskip 3pt}
        \hline 
    \end{tabular}
    \begin{tablenotes}
        \item Note: (1) Position where the converging ﬂows are measured in Figure \ref{fig:vel_grad}. (2) Filament length along which the velocity gradient is measured {(not corrected for inclination)}. (3) Filament mass within $L_{\rm ﬁl}$.  (4) Line mass of the filament within $L_{\rm ﬁl}$. (5) Velocity gradient measured along the ﬁlament. (6) Mass inflow rate of the ﬂow motion along the ﬁlament by a simple cylindrical model with $\tan\alpha$ = 1. (7) Mass inflow rate per unit length. ({8) Mass accretion rates assuming an SFE of 20$\%$. (9) Mass gain by the hub in $2\times10^{4}$ years with the adopted SFE.}
    \end{tablenotes}    
\end{center}
\end{table*}


\begin{figure*}
\centering
\includegraphics[width=\textwidth]{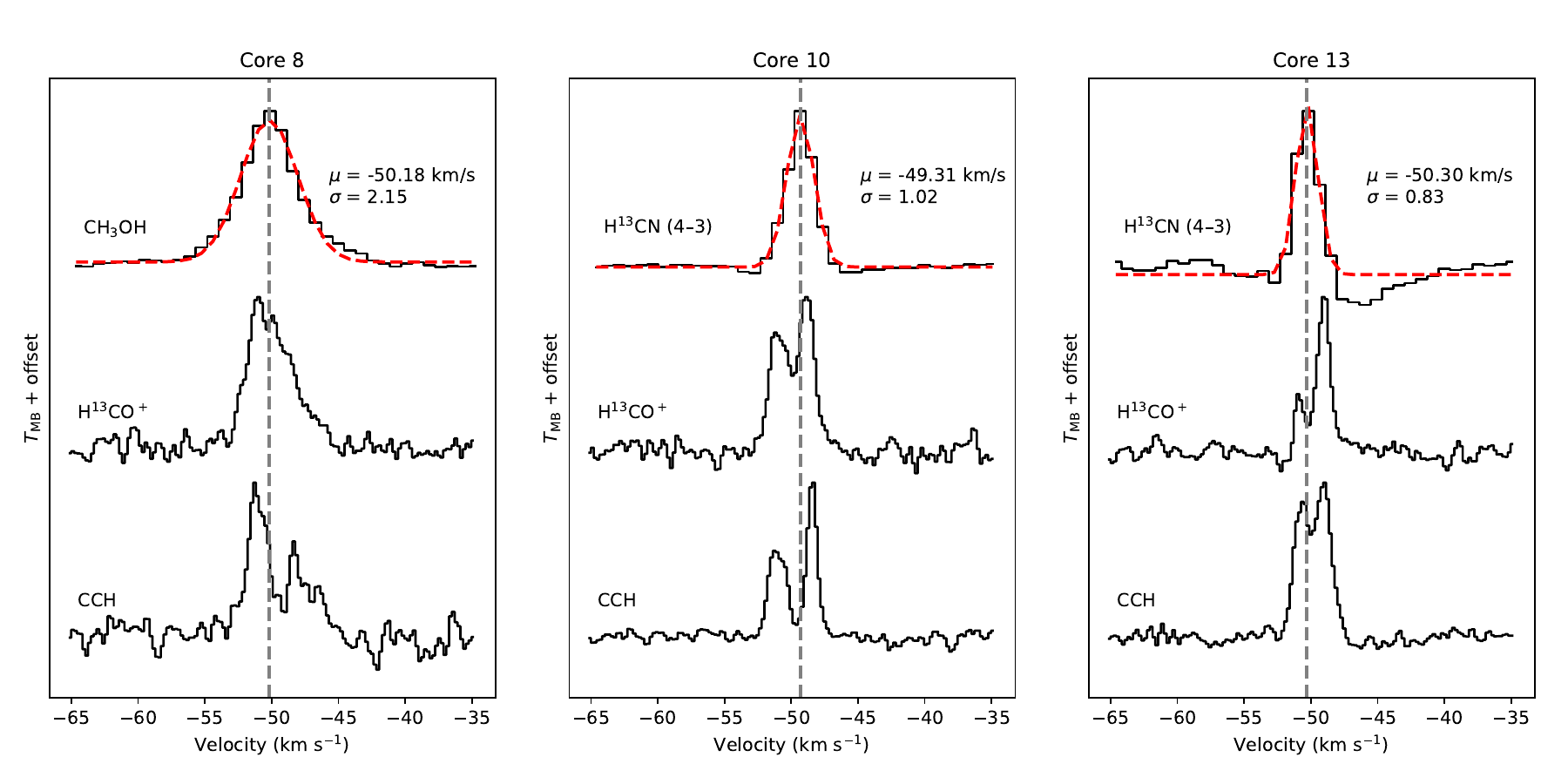}
\caption{Average spectra of H$^{13}$CO$^{+}$, CCH, and CH$_{3}$OH {(13$_{1,12}$--13$_{0,13}$)}for core 8 and H$^{13}$CN (4--3) for cores 10 and 13. The spectra of CH$_{3}$OH {(13$_{1,12}$--13$_{0,13}$)} and H$^{13}$CN (4--3) are fitted with a single Gaussian (red dashed line) to derive the centroid velocity ($\mu$) and velocity dispersion ($\sigma$). The centroid velocity corresponds to the systematic velocity (V$_{\rm lsr}$) of the cores, which is marked with vertical dashed lines in all three panels.}
\label{fig:infall_prove}
\end{figure*}


\begin{figure*}
\centering
\includegraphics[width=0.95\linewidth]{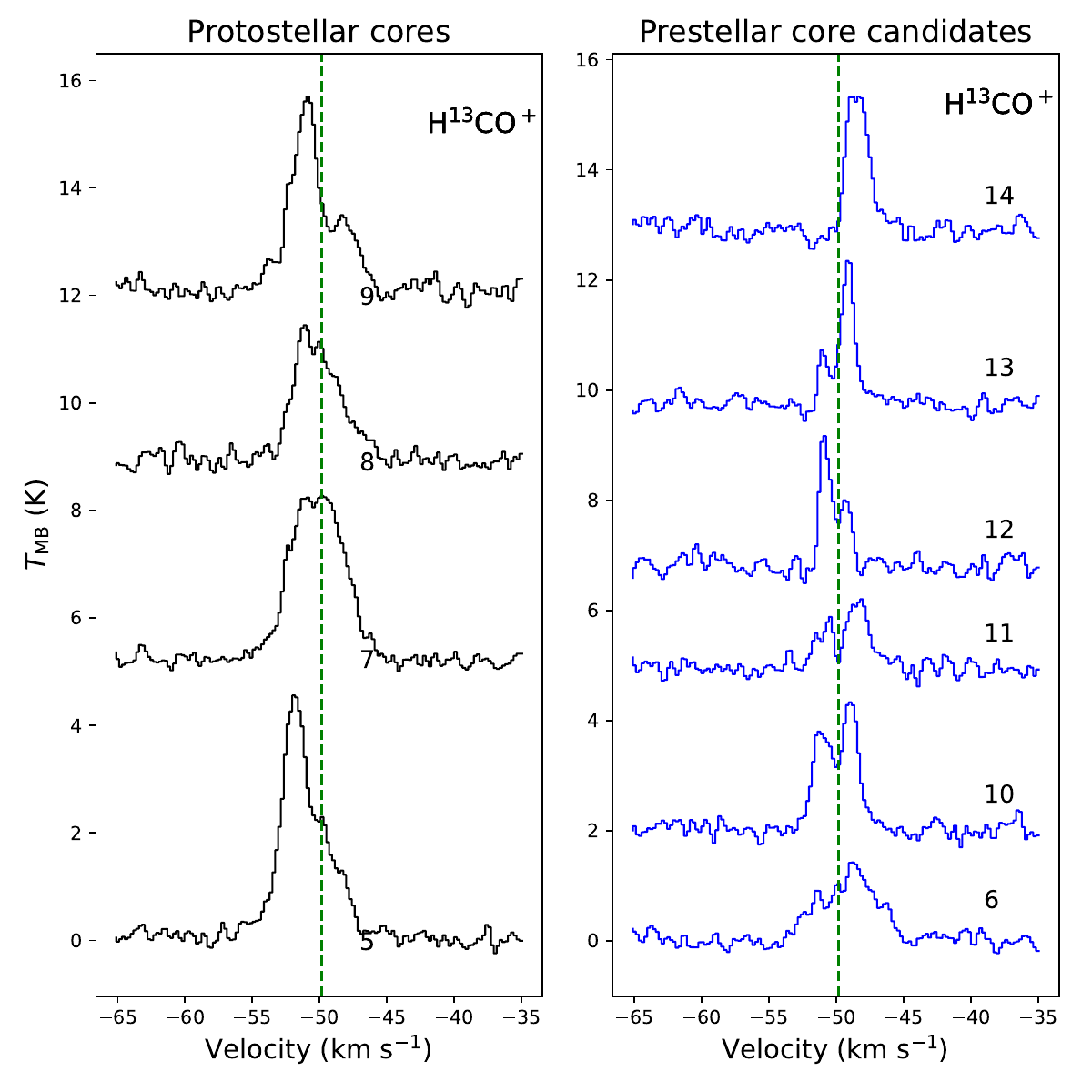}
\caption{H$^{13}$CO$^{+}$ spectra of the protostellar cores (left) and prestellar {core candidates} (right) in the region. The vertical green line shows the V$_{\rm{lsr}}$ of the protocluster.}
\label{fig:infall_H13COp}
\end{figure*}

\begin{figure*}
\centering
\includegraphics[width=\textwidth]{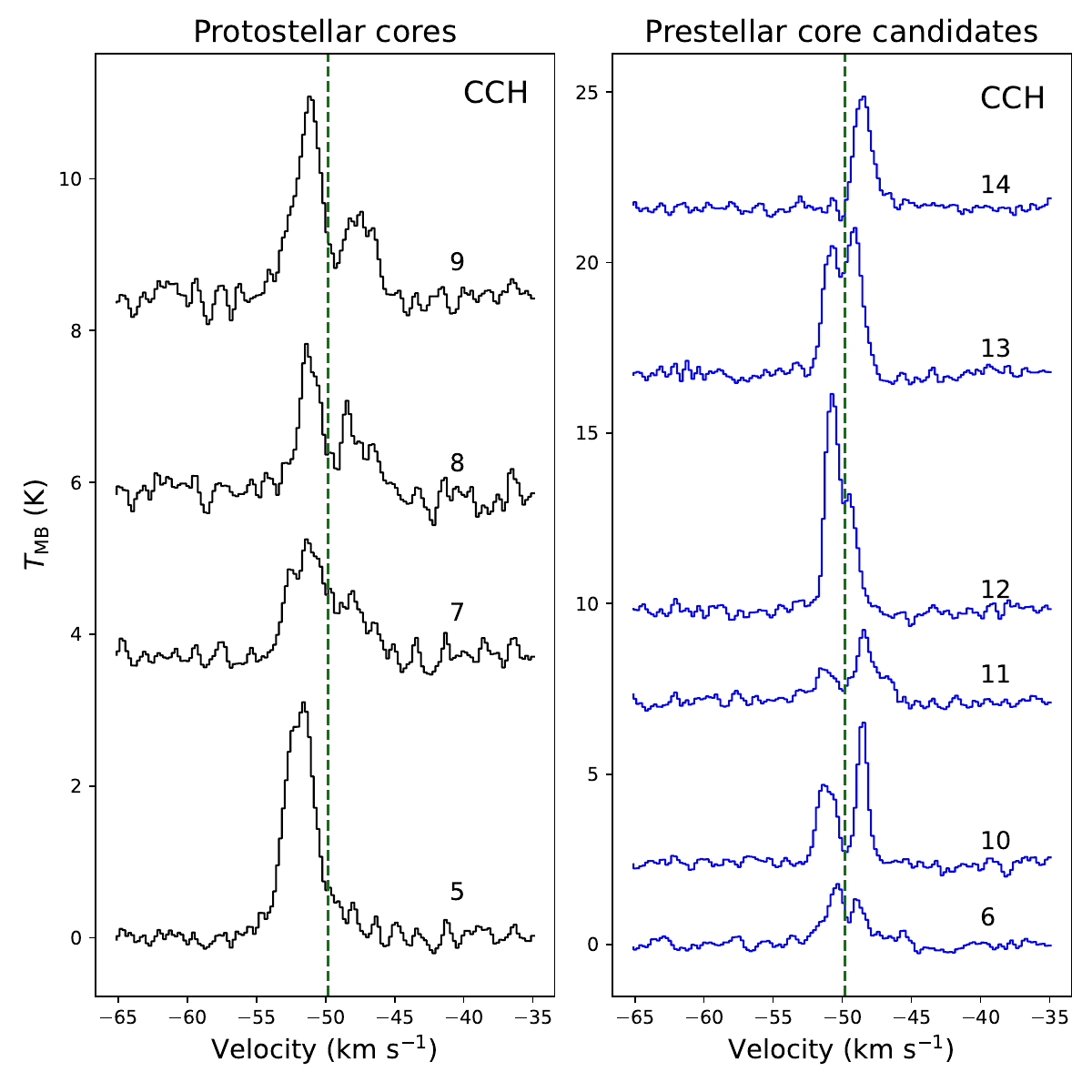}
\caption{Extracted CCH spectra of the protostellar {cores} (left) and prestellar {core candidates} (right). The vertical green line shows the V$_{\rm{lsr}}$ of the protocluster.}
\label{fig:infall_CCH}
\end{figure*}

\subsection{Gas infall rates in cores}
In previous section, we noted the inflow of gas along the filaments. It is now important to examine the gas dynamics of the cores present in this protocluster. For investigation of the potential collapse motions within the prestellar and protostellar cores, we analyze the spectrum of an optically thick tracer. Spectrum of such tracer is often useful to identify the characteristic self-absorption features with asymmetric double-peaked line profiles. A stronger blue peak in the double-peaked profile compared to the red one is commonly referred as blue profile and is typically considered as the signature of infalling gas. In contrary, a double-peaked profile with stronger red peak, known as red profile, signifies the expanding gas. 
Occasionally, in case of prestellar cores, red and blue profiles may signify infall and expansion motions, respectively {\citep{2002ApJ...577..798D}}. Such profile is generally expected when inversion of temperature gradient occurs from core to the envelope due to the presence of an external heating source \citep{2002ApJ...577..798D}, but this is not the case here.

In general, tracers like HCO$^{+}$ (1--0) or CS (2--1) are best-served for this purpose. However, in our protocluster, the HCO$^{+}$ line is significantly contaminated by outflow activities from the embedded protostars, while the CS line has a coarse spectral resolution of about 3 km s$^{-1}$ that can significantly distort these characteristic features.

Note that \cite{2022MNRAS.514.6038Z, 2023MNRAS.520..322Z}  have reported asymmetric and double-peaked profiles in H$^{13}$CO$^+$ towards the densest regions in several protoclusters of the ATOMS survey. 
Interestingly, we also found asymmetric blue H$^{13}$CO$^{+}$ profiles for most of the cores, with the V$_{\rm lsr}$ of the protocluster lying close to the dip between the blue and red parts, indicating the self-absorption feature, deviating from its commonly assumed optically thin nature in dense cores. {To further confirm this, we use optically thin tracers--CH$_{3}$OH (13$_{1,12}$--13$_{0,13}$) for the protostellar cores (5, 7, and 8), which is a tracer of hot cores, and H$^{13}$CN (4--3) for the prestellar core {candidates} (10 and 13). For other prestellar core {candidates}, we could not obtain the H$^{13}$CN (4--3) spectra might be because of lower flux level and/or significant missing flux. We fit Gaussians to the average spectra of these optically thin tracers to determine the V$_{\rm lsr}$ of each core. Figure \ref{fig:infall_prove} shows that the V$_{\rm lsr}$ of these tracers coincides with the dip in the H$^{13}$CO$^{+}$ profiles for cores 8, 10, and 13, further confirming that the double-peaked shapes arise from self-absorption and not from multi-component cloud. We also compare other infall tracers, such as CS, H$^{13}$CN (1–0), and CCH, for core 5  and 7 (Figure \ref{fig:core5}; Appendix C), which also clearly exhibits infall profiles. Hence, we infer that similar profiles in other cores where no spectrum was obtained for optically thin tracer, are also signifying the infalling gas.} We used H$^{13}$CO$^+$ line to analyze the infall motions for the cores. The  H$^{13}$CO$^{+}$ spectral profiles of protostellar {cores} and prestellar core {candidates} are separately shown in Figure~\ref{fig:infall_H13COp}. As can be seen in Figure~\ref{fig:infall_H13COp}, protostellar cores 5, 8, and 9 exhibit blue profiles. Among the prestellar {candidates}, cores 3, 12, 15, and 16 show blue profiles, while cores 10, 11, and 13 show red profiles suggestive of expanding gas.

\begin{figure*}
\centering
\includegraphics[width=\textwidth]{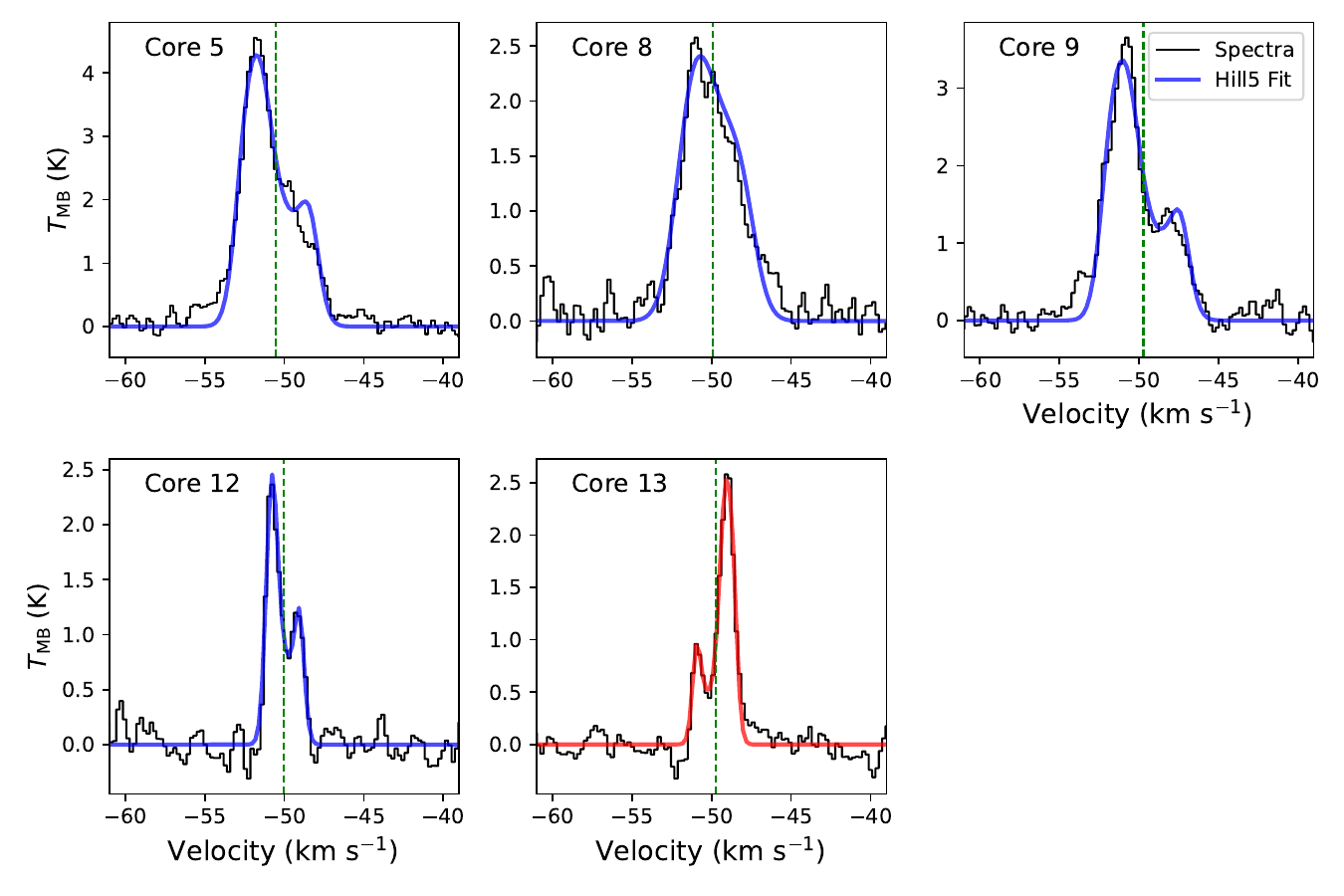}
\caption{Model fit to the H$^{13}$CO$^{+}$ spectral profiles (shown in black) of the cores. The profiles show the best-fit results with the Hill5 model \citep{2005ApJ...620..800D}. Model fitted spectra to the blue profiles are shown in blue color while the fit to the red profile is shown in red color. The green dashed lines shows the V$_{\rm lsr}$ of each core derived from HILL5.}
\label{fig:Hill5_fit_all}
\end{figure*}

\begin{figure*}
\centering
\includegraphics[scale=0.7]{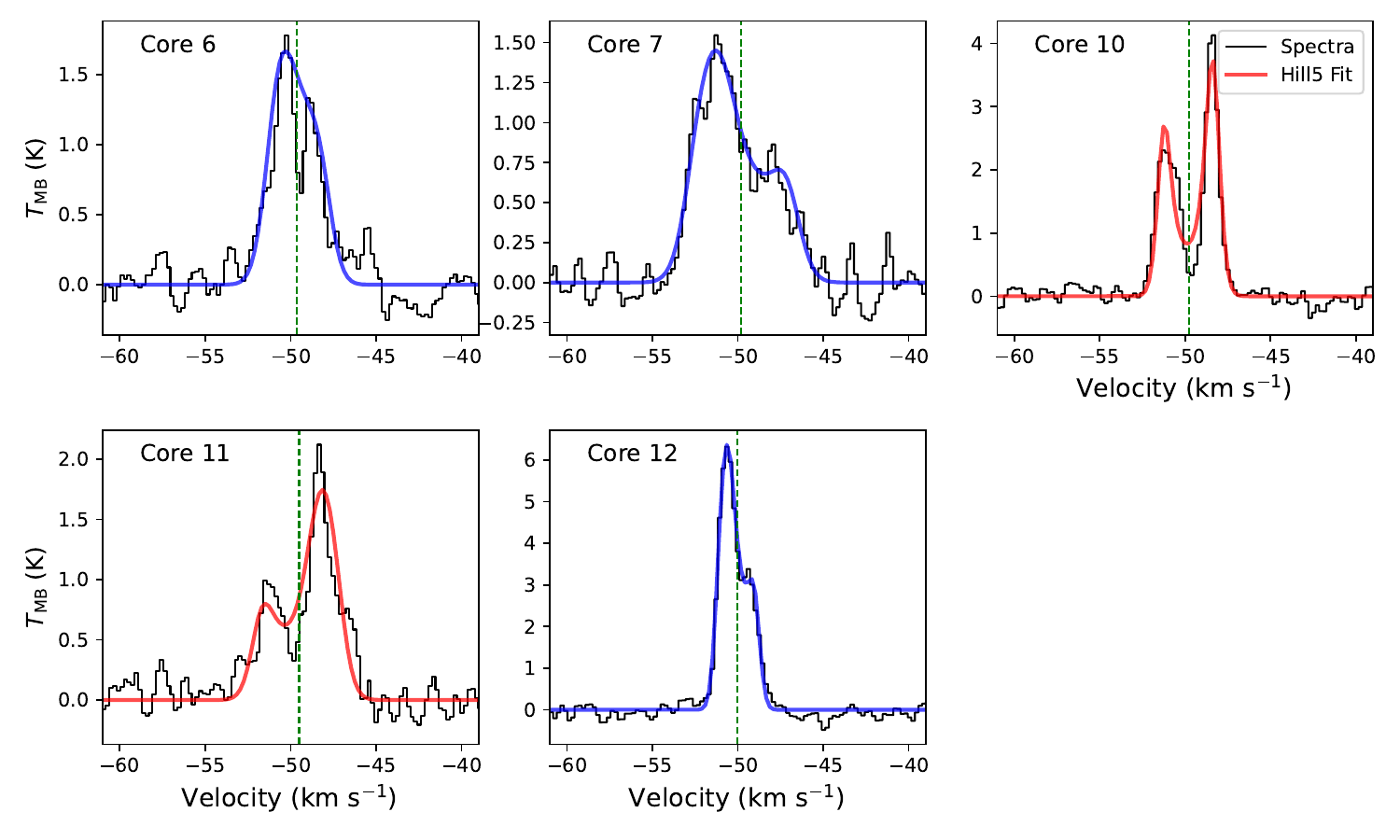}
\caption{Model fit to the CCH spectral profiles (shown in black) of the cores. The remaining symbols are the same as in the caption of Figure~\ref{fig:Hill5_fit_all}.}
\label{fig:Hill5_fit_CCH}
\end{figure*}

We also examined the CCH F(2–1) line profile (see Figure~\ref{fig:infall_CCH}), which is reported to be a good tracer for infalling gas motions in both prestellar and protostellar cores \citep{2009A&A...505.1199P, 2010ApJ...722.1633S}. {The line profile of CCH also revealed a similar behavior, i.e., the V$_{\rm lsr}$ obtained from the thin tracers coinciding with the dip in CCH (see Figure \ref{fig:infall_prove})}, with two additional cores showing blue profiles (core 7 in protostellar and core 6 in prestellar candidate). This may be attributed to the high optical thickness of H$^{13}$CO$^{+}$ in these regions, which can suppress the blue-shifted emission and thereby obscure typical infall signatures \citep{2000ApJ...538..260G}. The profiles for all the remaining cores show similar signatures in both H$^{13}$CO$^{+}$ and CCH spectral profiles.


After successful identification of blue profiles, we went on to estimate the infall velocity of the gas in these cores. The infall velocity of the cores with blue profile was determined by fitting the observed spectral profiles of H$^{13}$CO$^{+}$ and CCH with the Hill5 infall model \citep{2005ApJ...620..800D}. This model is developed for deriving infall velocities in a contracting molecular cloud by dealing with radiative transfer processes in two approaching layers whose excitation temperatures linearly increase toward the inner region \citep[see][for details]{2005ApJ...620..800D}. 
To reliably determine the infall velocity from the asymmetric line profiles using the Hill5 model, we applied a signal-to-noise (S/N) threshold $\gtrsim$ 15 to the observed spectra. The S/N ratio was calculated based on an rms noise level of 0.27 K, estimated from the line-free channels using \texttt{FUNStools}.\footnote{\href{https://github.com/radioshiny/funstools}{https://github.com/radioshiny/funstools}}

The overall fitting procedure is summarised in Appendix \ref{hill5} and fitting to the average profile of H$^{13}$CO$^{+}$ for core 9 is shown in Figure~\ref{fig:core9_cornerplot} as an example. The best-fit Hill5 profiles overlaid with the averaged line profiles of H$^{13}$CO$^{+}$ and CCH are shown in Figures \ref{fig:Hill5_fit_all} and \ref{fig:Hill5_fit_CCH}, respectively. In both figures, the red and blue color fits represent red and blue profiles, respectively. Fit to a few spectra (cores 7, 10, and 14 for H$^{13}$CO$^{+}$, and cores  5, 8, 9, 13, and 14 for CCH) were unreliable due to the complex nature of their spectra, and hence, they were excluded from further analysis. 

The resulting infall velocities (V$_{\rm in}$) are listed in Table \ref{tab:photometry}. By convention, positive V$_{\rm in}$  correspond to a blue profile, and negative V$_{\rm in}$ indicate a red profile. V$_{\rm in}$  vary from -0.73 to 1.24 km s$^{-1}$ for the cores (see columns 4 and 5 of Table \ref{tab:photometry}).
Note that for cores where V$_{\rm in}$ is derived for both the tracers (cores 10, 11 and 12), the obtained values are found to be consistent with each other.

The estimated infall velocities were further used to derive the mass infall rate ($\dot{M}_{\rm{in}}$) of gaseous material following the equation \citep [e.g.][]{2001ApJS..136..703L, 2022ApJ...940..112K, 2025arXiv251116978M}:
\begin{equation}
\dot{M}_{\rm{in}}=\frac{3M_{\rm in}}{R_{\rm in}}V_{\rm{in}}
\end{equation}
where $M_{\rm{in}}$ is the core mass, and $R_{\rm in}$ is the core radius, {adopted from \cite{2024ApJS..270....9X}}. The calculated values of $\dot{M}_{\rm{in}}$ are given in Table \ref{tab:photometry}.

We find that cores located in the hub region, namely cores 5, 7, 8, and 9, exhibit a high mass-infall rate, with values of 54.9, 196.2, 33.1, and 20.1 $\times$ 10$^{-5}$ M$_{\odot}$yr$^{-1}$, respectively. In contrast, cores 10 and 11 (with red profiles) show expansion of gas at a rate of 68.9 and {39.3}$\times10^{-5}$ M$_{\odot}$yr$^{-1}$, respectively. Other cores exhibit comparatively lower infall rates, consistent with their peripheral positions and possibly earlier evolutionary stages (see Section \ref{sec:infall-inflow} for more discussion).

\begin{table*}
\centering
    \caption{Estimated physical and dynamical parameters of the cores}
    \label{tab:photometry}
    \begin{tabular}{lcccccccc} 
        \hline
        Core ID & Mass & Radius & {V$_{\rm{lsr}}$} & V$_{\rm{in}} (\rm{H^{13}CO^{+}})$  & V$_{\rm{in}}(\rm{CCH})$& $\dot{M}_{\rm{in}}(\rm{H^{13}CO^{+}})$ & $\dot{M}_{\rm{in}}$(CCH)  \\
             &  (M$_{\odot}$) &  (AU) &  {(km s$^{-1}$)}   &  (km s$^{-1}$)  &  (km s$^{-1}$) & \rm ($\times$10$^{-5}$ M$_{\odot}$yr$^{-1}$) & \rm ($\times$10$^{-5}$ M$_{\odot}$yr$^{-1}$)& \\
           (1)   &  (2) &  (3)   &  (4)  &  (5) &  (6) &  (7) & (8) \\
        \hline
        5  & 4.0$\pm$2.4 & 4550 & {$-$50.61} &  1.05$^{+0.12}_{-0.11}$  & - & 54.9$\pm$35.1 & -  \\
        6  & 0.9$\pm$0.7 & 4840 & - & -& 0.56$^{+0.28}_{-0.31}$  & -& 6.6$\pm$5.1   \\
        7  & 9.7$\pm$5.2 & 3680 & {$-$50.70} &  - &  1.24$^{+0.26} _{-0.27}$ &- &196.2$\pm$110.9  \\
        8  & 1.0$\pm$0.6 & 1530 & {$-$50.18} &  0.80$^{+0.16}_{-0.18}$ & -&33.1$\pm$19.9&-   \\
        9  & 1.0$\pm$0.6 & 3740 & - & 1.08$^{+0.10}_{-0.10}$  & -& 20.1$\pm$11.0 &- \\
        10 & 6.0$\pm$4.1 & 3360 & {$-$49.31} & $-$0.16$^{+0.06}_{-0.07}$ &$-$0.14$^{+0.03}_{-0.03}$  & $-$68.9$\pm$41.7  & $-$46.3$\pm$31.7 \\
        11 & 1.3$\pm$1.0 & 1530 & - & $-$0.73$^{+0.32}_{-0.56}$& $-$0.73$^{+0.18}_{-0.32}$ & $-$39.3$\pm$30.2 & - \\
        12 & 1.1$\pm$0.8 & 3770 & - &  0.44$^{+0.08}_{-0.06}$ &  0.45$^{+0.11}_{-0.07}$&5.0$\pm$3.6&-  \\
        13 & 1.3$\pm$0.9 & 1530 & {$-$50.30} & $-$0.49$^{+0.08}_{-0.10}$ & -& $-$26.4$\pm$18.2 & -  \\
        14 & 5.5$\pm$3.5 & 3740 & - &  - & - &-&- \\
        \hline 
    \end{tabular}
    \begin{tablenotes}
        \item Note: (1) Core IDs. (2) \& (3) Mass and radius of the cores, respectively, as adopted from \cite{2024ApJS..270....9X}. (4) {V$_{\rm{lsr}}$ of each core derived from thin tracer  CH$_3$OH (13$_{1,12}$–13$_{0,13}$) and H$^{13}$CN (4–3). }(5) \& (6) Infall velocities of the cores derived using the Hill5 model from the average spectra of H$^{13}$CO$^{+}$ and CCH, respectively. (7) \& (8) Mass infall rates for each of the cores derived from V$_{\rm in}$(H$^{13}$CO$^{+}$) and V$_{\rm in}$(CCH), respectively{; here negative mass infall rates correspond to expansion.} (9) Infall time.
    \end{tablenotes}    
\end{table*}

\section{Discussion} \label{sec:discussion}
In Section~\ref{sec:result}, we found that there exists at least four filaments that show evidence of carrying gas to the central hub where multiple star-forming cores are present. Also, these filaments are inflowing gas at a rate of more than 10$^3$ M$_\odot$ Myr$^{-1}$. With such high gas inflow rate, these filaments are capable of transporting enough amount of gas to the central hub for the formation of massive stars. In fact, all the cores located at the central hub are showing the signature of infalling gas. In this section, we discuss our overall analyses in terms of the star formation activity in the protocluster. 


\subsection{M-R relationship of the cores}

\begin{figure}[!htb]
\centering
\includegraphics[width=1\linewidth]{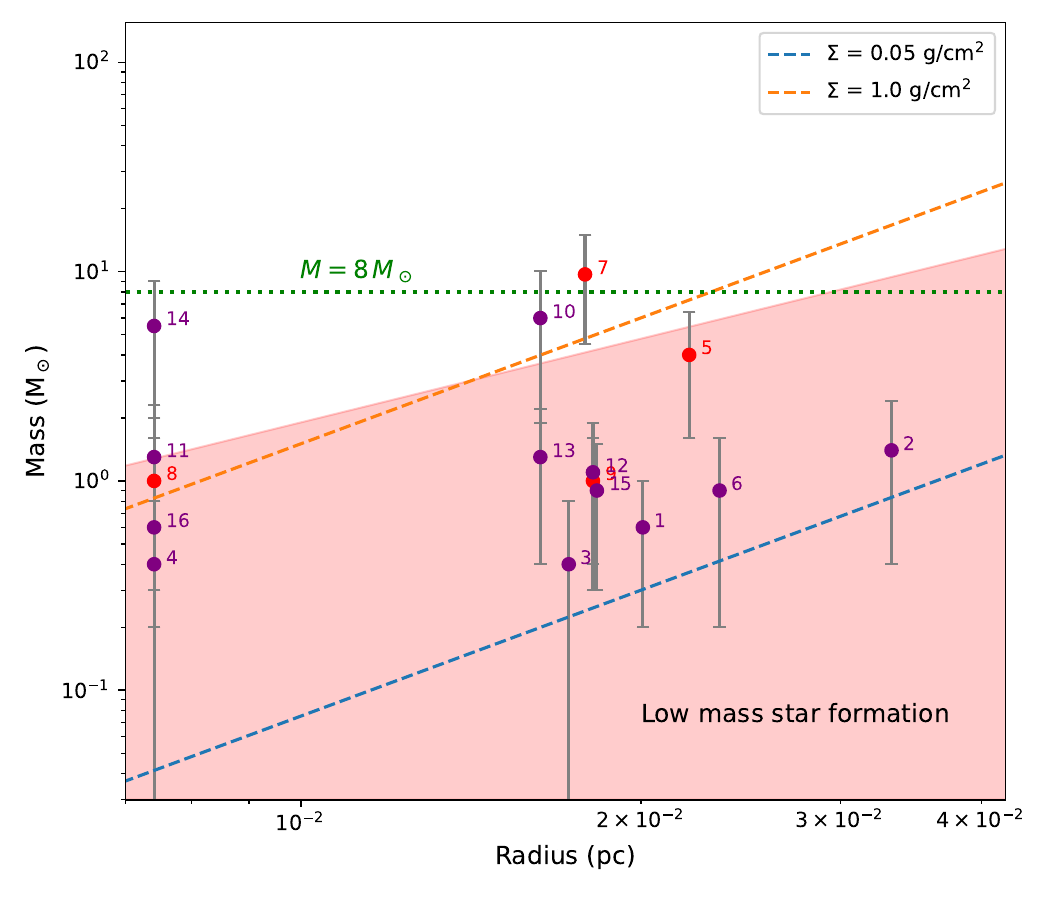}
\caption{The radii vs masses of the cores. The red circles mark the protostellar cores, while the purple circles represent the prestellar core {candidates}. The shaded region represents the area of low-mass star-forming cores that does not satisfy the criteria of m(r) $>$ 870\,M$_\odot$\,(r/$\mathrm{pc}$)$^{1.33}$. Surface density thresholds of 0.05 g cm$^{-2}$, and 1 g cm$^{-2}$ are shown as blue and orange dashed lines, respectively. The green dotted line is for a core mass of 8 M$_\odot$.}
\label{fig:massvsradius}
\end{figure}

To further examine the star-forming nature of the cores, we analysed their mass-radius relation that provides important clues whether a core will evolve into a massive star or into a low-mass star. The mass–radius relation of all the identified cores in this protocluster is shown in Figure~\ref{fig:massvsradius}. The shaded region represents the area of low-mass star-forming cores that does not satisfy the criteria of m(r) $>$ 870\,M$_\odot$\,(r/$\mathrm{pc}$)$^{1.33}$ \citep{2013ApJ...765L..35K}. It highlights the domain where low-mass cores are typically found, bounded below by the gravitational binding limit of $\Sigma=0.05$ g cm$^{-2}$ \citep{2008Natur.451.1082K}. Most of the cores lie below the critical surface density of $\Sigma=1.0$ g cm$^{-2}$ \citep{2014MNRAS.443.1555U}, suggesting they are associated with low-mass star formation.

The protostellar core, Core 7 residing in the  hub-region exceeds both the surface density threshold and the 8 M$_{\odot}$ mass limit, making it a potential candidate to evolve into a massive star. Note that cores 10, 11, and 14 also lie in the massive star formation regime but exhibit red profiles, indicating ongoing mass-loss or mass redistribution possibly to the central hub. To investigate this further, we look into the dynamical properties of the host filaments (F3 and F4) of the cores 10 and 11 as it is important to assess gas dynamics and understand the potential mass transfer towards the hub.


\begin{figure*}
    \centering
    \includegraphics[scale=0.7]{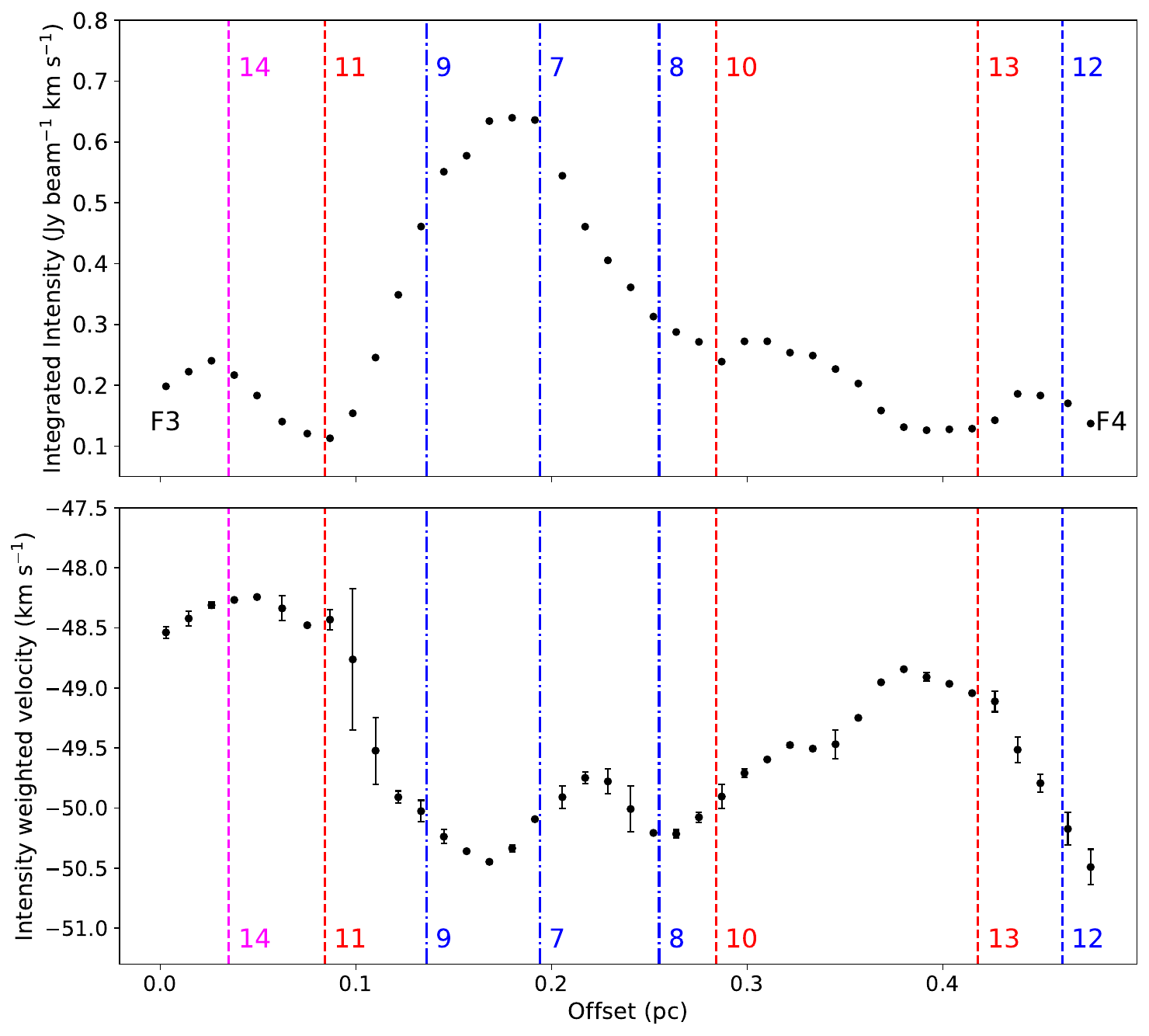}    
    \caption{The upper panel shows the distributions of integrated intensity along filament F3-F4, while the lower panel displays the intensity-weighted velocity (PV) along the same filament {using H$^{13}$CO$^+$}(see Figure \ref{fig:moment1}). {In both the panels, the values are uniformly sampled along the filament skeletons following the same methodology as described in Figure \ref{fig:vel_grad}. The velocity error bars represent the standard deviation of the velocities within each sampled bin along the offset.} 
    Blue and red vertical lines indicate the positions of dense cores with blue profile and red profiles, respectively, residing along the filament, whereas the magenta vertical line shows the position of the dense core with single-peaked profile. The dotted-dashed vertical blue lines at the central part shows the positions of the cores associated with the hub.}  \label{fig:velocity_gradient}
\end{figure*}

\subsection{Velocity Variation along the Filament }
Velocity and density fluctuations along filaments are noted in several studies \citep{2022ApJ...940..112K,2022MNRAS.514.6038Z}. Such fluctuations are likely caused by dense structures (i.e., cores) embedded within the filaments \citep{2019MNRAS.487.1259L}. Fluctuations along filaments may indicate oscillatory gas flows coupled to regularly spaced density enhancements that probably form via gravitational instabilities \citep{2020NatAs...4.1064H}. Figure \ref{fig:velocity_gradient} presents the intensity and the velocity distribution (i.e., PV) along the filaments in this protocluster including hub starting from F3 to F4. The velocity distribution exhibits a global minimum (lower panel), which spatially coincides with the peak of the integrated H$^{13}$CO$^{+}$ intensity (upper panel), corresponding to the center of the gravitational potential well (i.e., the hub-region). Clear velocity gradients can be noted on both sides of this hub region {(see Figure \ref{fig:vel_grad})} and the corresponding gradients were estimated in Section~\ref{sec:vel_grad}.

The positions of the cores are marked by vertical dashed lines in Figure \ref{fig:velocity_gradient}, with blue dotted-dashed lines representing the cores within the hub region, red and blue dashed lines denoting cores exhibiting red and blue-asymmetric spectral profiles, and pink corresponding to the rest of the cores that lie along filaments F3 and F4.
A clear {shallow dip} in the velocity profile is observed at the centroid position of Core 11, while Core 10 is located on a steep velocity gradient, {and both cores are located at minima in integrated intensity}, suggesting their dynamical connection with the hub region {i.e., Cores 10 and 11 are not gathering a significant amount of the mass}. Note that such a scenario is typically expected for competitive accretion theory, in which several cores/condensations together expected to create a global potential to transport gas from the outer part onto the gravitational center (i.e., hub), and feed the cores residing in the hub.

\subsection{Possible evolution of the cores}
The high velocity gradient and large mass inflow rates (in the order of 10$^4$ M$_\odot$ Myr$^{-1}$) towards the hub suggest that these filaments are actively feeding material into the central hub, where the protostellar cores reside. Such inflow can lead to rapid evolution of cores in the hub, potentially supporting massive star formation. A high infall rate exceeding $\sim$ 1000 M$_{\odot}$ Myr$^{-1}$ is typically required to form massive stars before radiative feedback halts further growth \citep{2003ApJ...585..850M}.

\cite{2002A&A...389..446D} estimate that the chemical timescale relevant for massive star formation ranges between $7 \times 10^3$ and $5 \times 10^4$ years, with a characteristic timescale of approximately $3 \times 10^4$ years. Similarly, \cite{2009Sci...323..754K} suggests that protostellar accretion proceeds smoothly over a period of roughly $2 \times 10^4$ years. Accumulation of a small fraction ($\sim$20-30\%) of inflowing gas (see Table \ref{tab:mass_aacretion}) onto the individual core during this timescale may lead to a significant mass growth and potentially lead to the formation of massive stars or a stellar cluster. This fractional conversion of inflowing mass into stellar mass is commonly quantified as the star formation efficiency (SFE). \cite{2000ApJ...545..364M} proposed SFE values in the range of 30–50\% for clustered low-mass star-forming regions and considered this range to represent upper limits for high-mass environments. Additionally, \cite{2007A&A...462L..17A} suggested a uniform SFE of 30\% $\pm$ 10\%, consistent with the observed shape of the stellar initial mass function (IMF).

For example, F3 has an inflow rate of 5135 M$_\odot$ Myr$^{-1}$. A core in the hub region with a typical 20\% accumulation efficiency of gas transportation rate of F3, can gain $\sim$20 M$_\odot$ in just 20,000 years \citep{2009Sci...323..754K}, which is substantial, especially when it already had a seed mass. {Columns 8 and 9 of Table \ref{tab:mass_aacretion}} list the mass inflow rates of the filaments (see Section \ref{sec:vel_grad}), a value of fractional mass inflow rate to the core assuming an efficiency of 20\%, and an estimation of the core mass accumulate over a timescale of $2 \times 10^4$ years with this effective mass inflow.

With a typical conjecture of mass inflow through the filaments $2 \times 10^4$ years at a constant efficiency of 20\%, the additional mass accreted by the hub would rise by 35 M$_\odot$, while it will increase by 52.5 M$_\odot$ if an efficiency of 30\% is assumed. In both cases, the accumulated mass is sufficient to form  massive stars within the hub.


\subsection{Infall and inflow rates: an overview} \label{sec:infall-inflow}
The global infall rate of (1.56$\pm$0.03) $\times$ 10$^{4}$ M$_\odot$ Myr$^{-1}$, estimated at a radius of 0.74 pc using the CO(4-3) line by \citet{2021RAA....21...14Y}, provides a large-scale view of accretion in this system. Zooming into a smaller scale, around 0.3 pc, we observe the individual filamentary structures feeding the central hub.

For filament F4, inflow rate is estimated to be {1297} M$_\odot$ Myr$^{-1}$. Core 12, located at the periphery of this filament, shows a small infall rate of $\sim$50 M$_\odot$ Myr$^{-1}$ ($\sim$3\% of the total inflow). Notably, as we move towards the hub-region along the filament, core 13 and  core 10, appear to be losing mass (with a negative infall rate of  $\sim$260 M$_\odot$ Myr$^{-1}$ and  $\sim$500 M$_\odot$ Myr$^{-1}$, respectively), possibly contributing material to the central hub (Core 8). This increasing negative infall rates is suggestive of increasing gravitational potential toward the hub. Within the hub region, core 8 exhibits the highest infall rate of 330 M$_\odot$ Myr$^{-1}$, accounting for $\sim$25\% of the total inflow—highlighting significant mass accumulation at the hub. 

For filament F3, the estimated inflow rate is {5017} M$_\odot$ Myr$^{-1}$. Core 14 shows no significant accretion activity. Moving towards the hub along the filament, core 11 appears to be losing mass at a rate of approximately 400 M$_\odot$ Myr$^{-1}$, likely feeding core 9, which itself is accreting at an infall rate of 200 M$_\odot$ Myr$^{-1}$. The contribution from the filament F3 is approximately {4}\% to the core 9. For filament F1, the inflow rate is estimated to be {1762} M$_\odot$ Myr$^{-1}$. Core 5 is collapsing with an infall rate of 550 M$_\odot$ Myr$^{-1}$, which corresponds to approximately {31}\% of the mass being supplied by the filament. For core 7 has the highest infall rate of 1962 M$_\odot$ Myr$^{-1}$. With all the three filaments feeding material to this core together, this corresponds to {25}\% of the combined inflow rate. It shows that in the hub region the infall rate is very high ({4}\% to {27}\%) to the total inflow rate of their respective filaments. Note that all these calculations are done in the immediate vicinity of the adjacent filament.

These results collectively demonstrate a hierarchical mass accretion structure: at larger scales (R $\sim$ 0.74 pc), the accretion rate is on the order of 10$^4$M$_\odot$ Myr$^{-1}$; at filament scales ($\sim$0.3 pc), it drops to 10$^{3}$M$_\odot$ Myr$^{-1}$; and at core scales, further down to 10$^{2}$ M$_\odot$ Myr$^{-1}$. This progression strongly supports the hub–filament paradigm, where gravitational potential wells at the hub center govern the mass flow. The increasing inflow toward the hub and the transition from low infall in peripheral cores to enhanced accretion in inner cores illustrate the dynamic, gravity-driven nature of star-forming environments.


\section{Summary} \label{sec:summary}
We investigated the massive protocluster G318.049+00.086, utilizing molecular line and continuum data from the ATOMS, ASSEMBLE and QUARKS surveys. This region has been identified as a hub-filament system. Recently, \citet{2024ApJS..270....9X} reported the presence of 16 cores within this protocluster, comprising 4 protostellar and 12 prestellar core {candidates}, with the protostellar ones concentrated in the hub region.

Using H$^{13}$CO$^{+}$ and CCH spectral line data, along with 1.3 mm continuum emission, we explored the gas dynamics of the embedded cores, and the role of filamentary structures in the formation and evolution of these cores. Our main findings are summarized below:

\begin{itemize}
    \item  The 1.3 mm continuum with higher spatial resolution reveals four filaments in the densest regions covering the hub and its vicinity. These filaments (named as F1-F4) spatially coincide with the cores near the hub-region.
    \item The H$_2$ column density map was derived from the N(H$^{13}$CO$^{+}$) column density map. While the estimated filament masses are typical for Galactic filaments, about 100 M$_\odot$, the estimated line masses of the filaments are found to be 992-1746 M$_\odot$ pc$^{-1}$ which are typical for giant filaments.
    \item {The velocity gradients along filaments F1, F3, and F4 were 17.8 km s$^{-1}$, 22.1 km s$^{-1}$ and 10.9 km s$^{-1}$, respectively, and corresponding mass inflow rates were 1762 M$_\odot$ Myr$^{-1}$, 5017 M$_\odot$ Myr$^{-1}$, and 1297 M$_\odot$ yr$^{-1}$. }
    \item The cores associated with the protocluster show blue- and red-asymmetric line profiles in both H$^{13}$CO$^{+}$ and CCH spectra, indicative of infalling and expanding gas. Interestingly, majority of the protostellar cores show blue profiles, signature of infalling gas. A few prestellar core {candidates} located on the filaments showed red profiles. Infall velocities were determined using Hill5 model fits with MCMC sampling, yielding mass infall rates of approximately $\sim$(7-196)$\times$10$^{-5}$  M$_\odot$ Myr$^{-1}$.
    \item The mass–radius (M–R) relationship suggests that while most cores lie in the low-mass star formation regime, those situated in the high-mass regime predominantly show red profiles. This scenario possibly signifying that although they have the potential to form a massive star, they are significantly losing their mass either in form of expansion or to feed the central hub. In fact, PV analysis confirms that two of these cores are funnelling mass into the hub region, supporting the competitive accretion scenario.
    \item The mass inflow rates along the filaments are consistently higher than the core-level infall rates, suggesting an ongoing and efficient mass supply mechanism from large to small scales. If the filamentary accretion continues uninterrupted, the central hub is expected to accumulate sufficient mass to support the formation of massive stars. 
\end{itemize}

\section{Acknowledgments}
{We thank the anonymous referee for the constructive comments and suggestions.} G.G. gratefully acknowledges support by the ANID BASAL project FB210003. AH thanks the support by the S. N. Bose National Centre for Basic Sciences under the Department of Science and Technology, Govt. of India and the CSIR-HRDG, Govt. of India  for the funding of the fellowship. CWL is supported by the Basic Science Research Program through the National Research Foundation of Korea (NRF) funded by the Ministry of Education, Science and Technology (NRF-2019R1A2C1010851), and by the Korea Astronomy and Space Science Institute grant funded by the Korea government (MSIT) (Project No. 2022-1-840-05).



\appendix

\section{Moment 0 map of all line tracer in ATOMS} \label{moment_map}

\setcounter{figure}{0} \renewcommand{\thefigure}{A.\arabic{figure}}
\setcounter{table}{0} \renewcommand{\thetable}{A.\arabic{table}}



\begin{figure*}
\centering
\includegraphics[width=\textwidth]{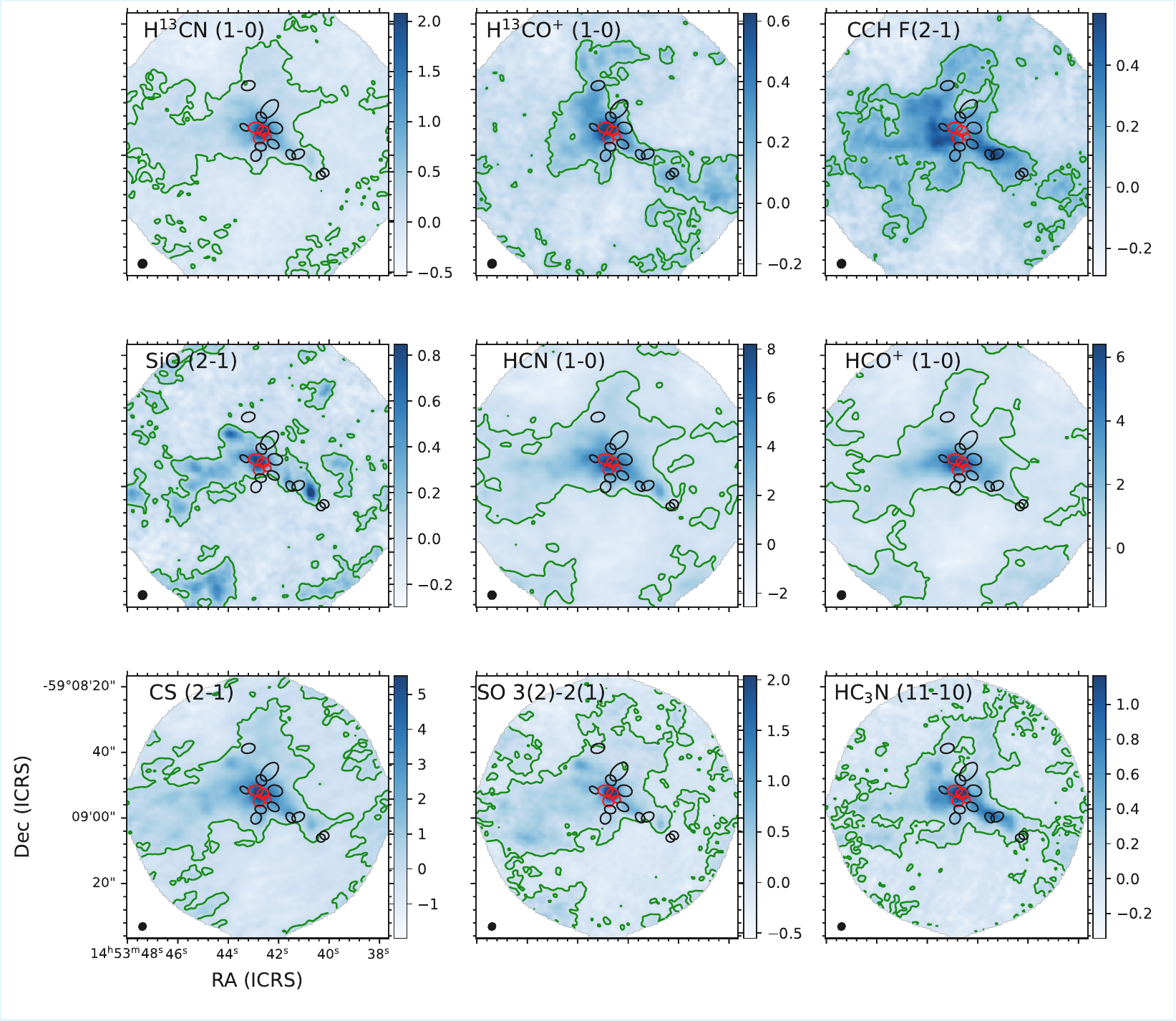}
\caption{{Moment 0 maps of the molecular transitions observed towards G318.049+00.086. The beam sizes are shown in black at the left corner of each subplot. The color bars indicate integrated intensities in units of Jy beam$^{-1}$ km s$^{-1}$. The ellipses represent the adopted cores at 870 $\mu$m by \cite{2024ApJS..270....9X}. Red and black ellipses denote protostellar cores and prestellar core {candidates}, respectively. Green contours are drawn at the 5 $\sigma$  level. The  $\sigma$  values for the individual transitions are 10.2, 11.1, 11.7, 10.9, 15.7, 16.1, 4.8, 5.1, and 5.2 Jy beam$^{-1}$ km s$^{-1}$, respectively, following the panel order.}}
\label{fig:moment0_all}
\end{figure*}

To show the comparison between all the molecular lines, we have over-plotted all the cores on the velocity-integrated intensity map (see Figure \ref{fig:moment0_all}). 

The moment 0 maps reveal that many of the molecular tracers, including HCN, HCO$^+$, CS, and CH$_3$OH, peak toward the core positions, showing strong spatial correlation with the 3 mm continuum. This suggests that these lines are closely associated with ongoing star formation activity. In contrast, tracers like CS, SO and SiO showing the spatial correlation by showing shock region and tracing same nodes. HCN and HCO$^{+}$ are showing similar correlation and tracing the outflow activity. H$^{13}$CN, H$^{13}$CO$^{+}$ and CCH F(2-1) are showing spatial correlation and tracing the same denser region. 


\section{Hill5 Model Fit to the Infall Profiles}\label{hill5}
The Hill5 model is a simple radiative transfer model used for deriving infall velocity in a contracting molecular cloud by dealing with radiative transfer processes in two approaching layers whose excitation temperatures linearly increase toward the inner region \citep[see Equation 9 of][]{2005ApJ...620..800D}. The model consists of a core with a peak excitation temperature ($T_{\rm peak}$) at the center and an excitation temperature of $T_{0}=0$ at the near and far edges of the core. The optical depth of the core is $\tau$, and its infall velocity is $V_{\rm infall}$, while the systematic velocity of the system is $V_{\rm LSR}$, and both the regions are assumed to have equal velocity dispersion of $\sigma$ for the observed molecule. These five parameters were set as free with initial guesses and bounds to fit the observed spectra with the model.

\begin{figure}
\centering
\includegraphics[width=\textwidth]{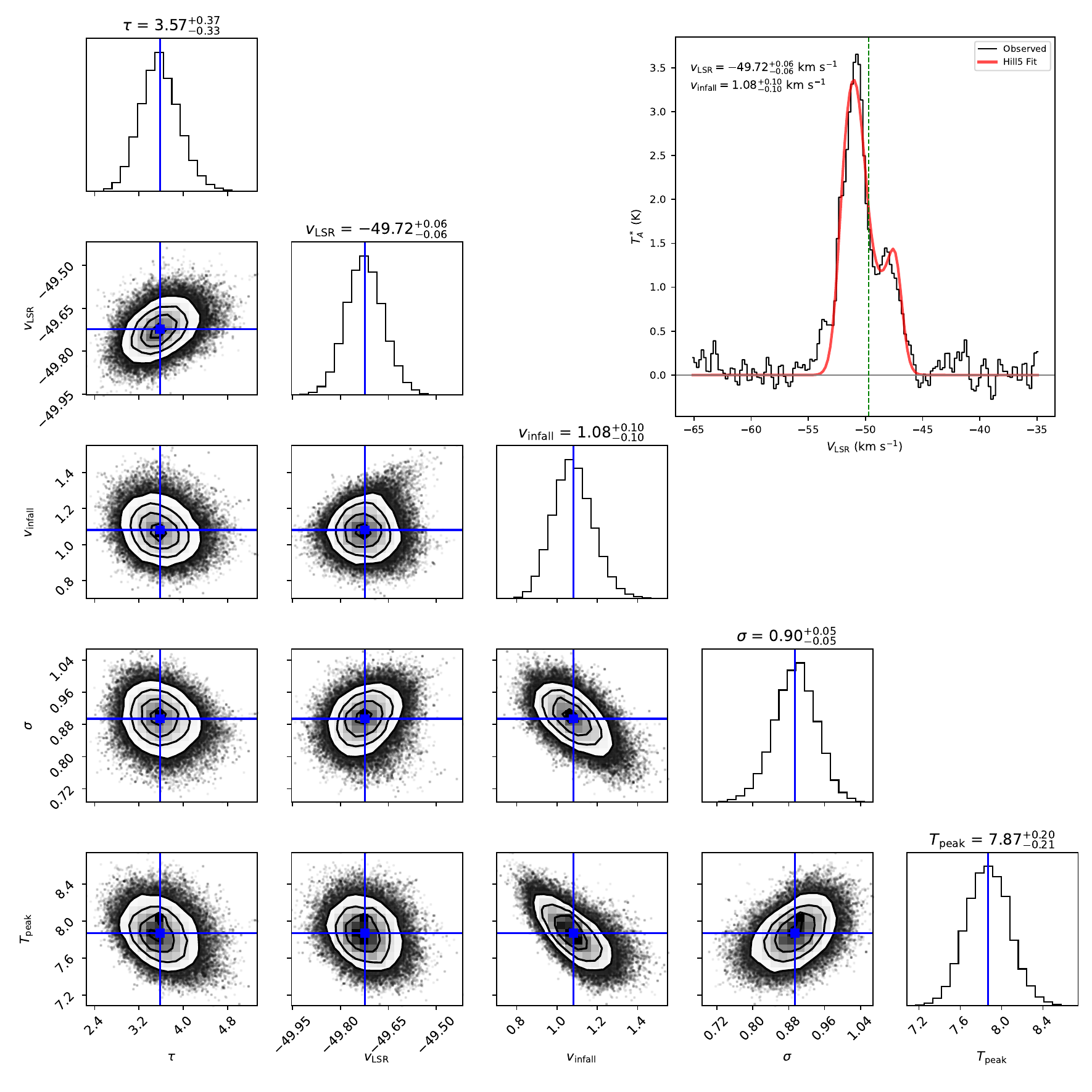}
\caption{Determination of infall velocity and local standard of rest velocity of core 9. The corner plots show the posterior probability distributions of parameters. The blue lines show the median values of the parameters. The best-fit model (red line) is over-plotted on the average spectra (black line) at the upper-right corner of the figure. The vertical green dashed line marks the obtained V$_{\rm lsr}$ from the fit.}
\label{fig:core9_cornerplot}
\end{figure}

For robust fitting with accurate estimation of the error of fitted parameters, the fitting was performed by Markov Chain Monte Carlo (MCMC) sampling using the \texttt{emcee}\footnote{\url{https://github.com/dfm/emcee}} Python package \citep{2013ascl.soft03002F}. The corner plots were generated to check the posterior probability distributions of parameters for each spectrum. In the initial trial, we set the parameter range very broad to identify the most probable region across possible local minima. The second trial uses this result to set more reasonable bounds on the parameters for a final fit. Figure \ref{fig:core9_cornerplot} shows the posterior distribution of the parameters in the second trial as well as the final fit to the spectra of core 9.

\section{Spectral profiles of core 5 and core 7}\label{spectra}

\begin{figure}[h]
\centering 
\includegraphics[scale=0.7]{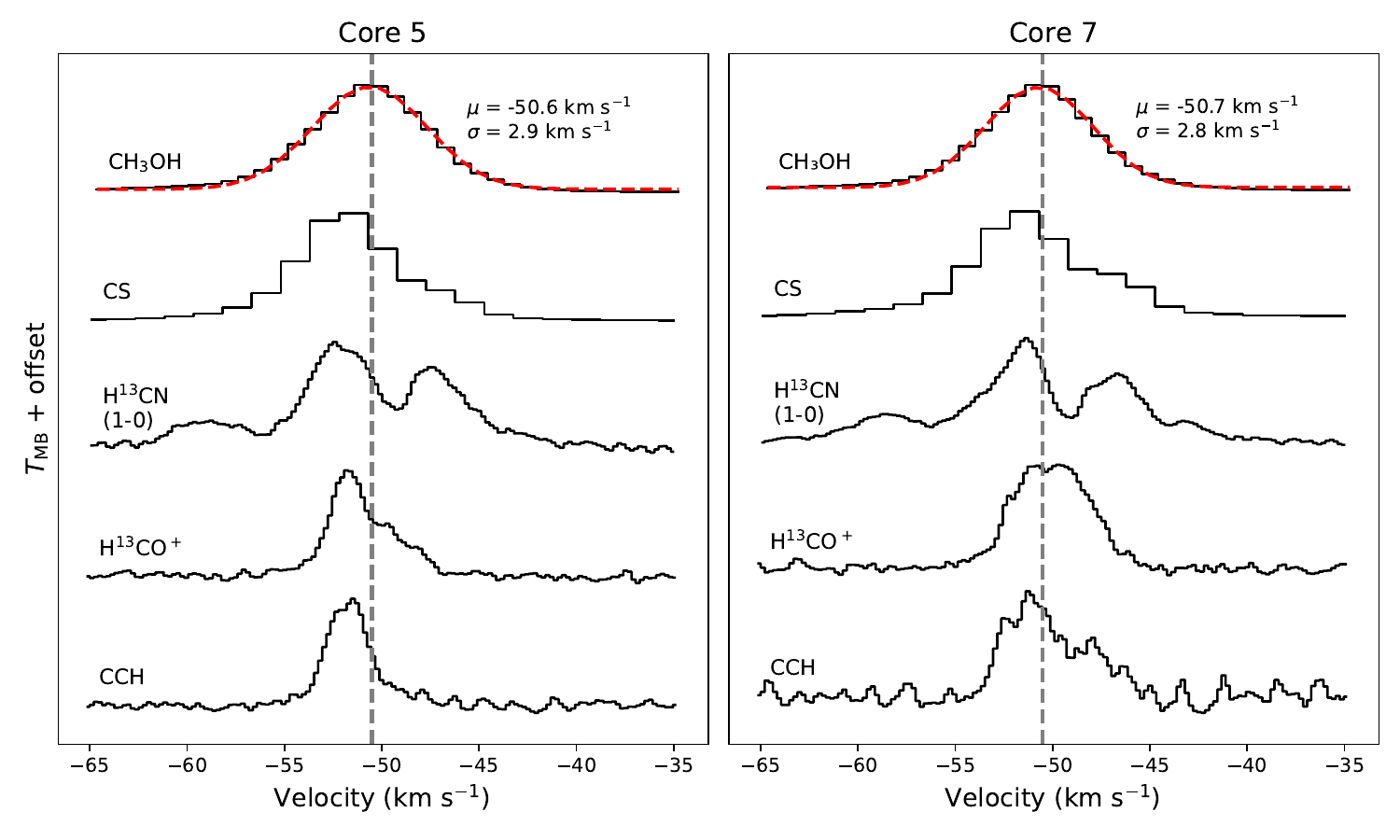}
\caption{{The average spectra of CH$_3$OH (13$_{1,12}$–13$_{0,13}$) and other infall tracers for cores 5 (left) and 7 (right). The CH$_3$OH (13$_{1,12}$–13$_{0,13}$) line is fitted with a single Gaussian as shown in red color. The fitted line centroid velocity ($\mu$), i.e., V$_{\rm lsr}$ of the core, and velocity dispersion ($\sigma$) are shown on the right. The vertical grey dashed line marks the obtained V$_{\rm lsr}$ from the fit.}}
\label{fig:core5}
\end{figure}

Figure \ref{fig:core5} presents the averaged spectral profiles of various molecular tracers for core 5 (left) and core 7 (right). The optically thin CH$_3$OH (13$_{1,12}$--13$_{0,13}$) line is fitted with a single Gaussian profile to determine the centroid velocity ($\mu$) and velocity dispersion ($\sigma$), where $\mu$ represents the systemic velocity (V$_{\rm lsr}$) of the core. Comparison with other tracers, including H$^{13}$CO$^{+}$, CS, H$^{13}$CN (1--0), and CCH, reveals clear evidence of infall signatures in both cores. The H$^{13}$CN (1--0) transition exhibits three hyperfine components, with the main component having an offset from V$_{\rm lsr}$ also shows signature of infall in both. For core 5, the derived V$_{\rm lsr}$ lies near the self-absorption dip between the blue and red peaks of H$^{13}$CO$^{+}$  and CS line profiles, indicating the ongoing infall motions, while the CCH emission also exhibits a similar displacement between its peak intensity and V$_{\rm LSR}$. These characteristics collectively suggest that the observed line asymmetries are primarily caused by infall motions rather than the presence of multiple velocity components along the line-of-sight. For core 7, located in the most dense hub region, a shallow dip is seen in the H$^{13}$CO$^{+}$ profile, while CCH shows a more prominent blue asymmetry. Although the CCH dip does not coincide exactly with V$_{\rm lsr}$, it is offset from the line peak; considering all tracers together, we interpret this as indicative of infall activity in the core.

\clearpage

\bibliography{sample631}{}
\bibliographystyle{aasjournal}

\end{document}